\documentclass[prl,twocolumn,letterpaper, superscriptaddress]{revtex4-1}
\usepackage{graphicx}
\usepackage{dcolumn}
\usepackage{bm}
\usepackage{color}
\usepackage{tabularx}
\usepackage{array}
\usepackage{amsmath}
\usepackage{stmaryrd}  
\usepackage{ulem}

\begin{document}

\title{Phase-resolved electrical detection of coherently coupled magnonic devices}

\author{Yi Li$^*$}
\affiliation{Materials Science Division, Argonne National Laboratory, Argonne, IL 60439, USA}
\thanks{Yi Li and Chenbo Zhao contributed equally to this paper}

\author{Chenbo Zhao$^*$}
\affiliation{Materials Science Division, Argonne National Laboratory, Argonne, IL 60439, USA}

\author{Vivek P. Amin}
\affiliation{Department of Chemistry and Biochemistry, University of Maryland, College Park, Maryland 20742, USA}
\affiliation{Physical Measurement Laboratory, National Institute of Standards and Technology, Gaithersburg, Maryland 20899, USA}

\author{Zhizhi Zhang}
\affiliation{Materials Science Division, Argonne National Laboratory, Argonne, IL 60439, USA}

\author{Michael Vogel}
\affiliation{Materials Science Division, Argonne National Laboratory, Argonne, IL 60439, USA}
\affiliation{Institute of Physics and Center for Interdisciplinary Nanostructure Science and Technology (CINSaT), University of Kassel, Heinrich-Plett-Strasse 40, Kassel 34132, Germany}

\author{Yuzan Xiong}
\affiliation{Department of Physics, Oakland University, Rochester, MI 48309, USA}
\affiliation{Materials Science Division, Argonne National Laboratory, Argonne, IL 60439, USA}

\author{Joseph Sklenar}
\affiliation{Department of Physics and Astronomy, Wayne State University, Detroit, Michigan 48202, USA}

\author{Ralu Divan}
\affiliation{Center for Nanoscale Materials, Argonne National Laboratory, Argonne, IL 60439, USA}

\author{John Pearson}
\affiliation{Materials Science Division, Argonne National Laboratory, Argonne, IL 60439, USA}

\author{Mark D. Stiles}
\affiliation{Physical Measurement Laboratory, National Institute of Standards and Technology, Gaithersburg, Maryland 20899, USA}

\author{Wei Zhang}
\affiliation{Department of Physics, Oakland University, Rochester, MI 48309, USA}
\affiliation{Materials Science Division, Argonne National Laboratory, Argonne, IL 60439, USA}

\author{Axel Hoffmann}
\affiliation{Department of Materials Science and Engineering, University of Illinois at Urbana-Champaign Urbana, IL 61801, USA}

\author{Valentyn Novosad}
\email{novosad@anl.gov}
\affiliation{Materials Science Division, Argonne National Laboratory, Argonne, IL 60439, USA}

\date{\today}

\begin{abstract}

We demonstrate the electrical detection of magnon-magnon hybrid dynamics in yttrium iron garnet/permalloy (YIG/Py) thin film bilayer devices. Direct microwave current injection through the conductive Py layer excites the hybrid dynamics consisting of the uniform mode of Py and the first standing spin wave ($n=1$) mode of YIG, which are coupled via interfacial exchange. Both the two hybrid modes, with Py or YIG dominated excitations, can be detected via the spin rectification signals from the conductive Py layer, providing phase resolution of the coupled dynamics. The phase characterization is also applied to a nonlocally excited Py device, revealing the additional phase shift due to the perpendicular Oersted field. Our results provide a device platform for exploring hybrid magnonic dynamics and probing their phases, which are crucial for implementing coherent information processing with magnon excitations.

\end{abstract}

\maketitle

Hybrid magnonic systems have recently attracted wide attention due to their rich physics and application in coherent information processing \cite{HueblPRL2013,TabuchiPRL2014,ZhangPRL2014,GoryachevPRApplied2014,BhoiJAP2014,TabuchiScience2015,ZhangNComm2015,BaiPRL2015,LachanceScienceAdvan2017,BaiPRL2017,HarderPRL2018,LiPRL2019_magnon,HouPRL2019,McKenziePRB2019,LachanceQuirionScience2020}. The introduction of magnons has greatly enhanced the tunability in hybrid dynamics, the capability of coupling to diverse excitations for coherent transduction \cite{OsadaPRL2016,ZhangPRL2016,ZhangScienceAdv2016,AnPRB2020,ZhaoChenboPRApplied2020}, as well as the potential for on-chip integration \cite{LiPRL2019_magnon,HouPRL2019,McKenziePRB2019}. Recently, thin-film-based magnon-magnon hybrid systems have provided a unique platform for implementing hybrid magnonic systems \cite{KlinglerPRL2018,ChenPRL2018,QinSREP2018,LiPRL2020_YIGPy,FanPRApplied2020,XiongSciRep2020,XiongIEEE2020,MacNeillPRL2019,LiensbergerPRL2019,DaiJAP2020}. Coupling between materials in the hybrid structure can arise through the interfacial exchange interaction. Because magnon excitations are confined within the magnetic media, it is convenient to build up more compact micron-scale hybrid platforms compared with millimeter-scale microwave circuits. Furthermore, abundant spintronic phenomena, such as spin-torque manipulation and spin pumping, can be used to control and engineer the hybrid dynamics especially for magnetic thin-film devices.

One important aspect of hybrid magnonic systems is controlling and engineering the phase relation between different dynamic components, leading to phenomena such as exceptional points \cite{ZhangDengkeNComm2017,ZhangPRL2019}, level attraction \cite{BhoiPRB2019,BoventerPRResearch2020} and nonreciprocity \cite{WangPRL2019,ZhangPRApplied2020} in cavity spintronics. Phase resolved detection of individual magnetization dynamics has been extensively explored electrically, optically, and with advanced light sources. In particular, electrical measurements of magnetic thin-film devices via spin rectification effects \cite{JuretschkeJAP1960,TulapurkarNature2005, FuchsAPL2007,GuiPRL2007,SankeyNphys2008,KubotaNphys2008} can directly transform microwave magnetic excitations into sizable dc voltages. This technique has been used to sensitively measure nanoscale magnetic devices and, more importantly, the phase of magnetization dynamics in order to quantify the spin torque generated from charge current flow \cite{ChenAPL2008,FangNNano2011,LiuPRL2011,BaiPRL2013,GuiSChinaPMA2013,MellnikNature2014,ZhangWeiPRL2014,SanchezPRL2014,SklenarPRB2015,NanPRB2015,JungfleischPRL2016,HarderPhysRep2016}.

In this work, we establish the usefulness of electrical excitation and detection for the study of coherently coupled magnon-magnon hybrid modes in Y$_3$Fe$_5$O$_{12}$/Ni$_{80}$Fe$_{20}$ (YIG/Py) thin-film bilayer devices. This approach differs from the previous work on inductive microwave measurements \cite{KlinglerPRL2018,QinSREP2018,LiPRL2020_YIGPy} in its applicability to nanoscale devices and its phase sensitivity. The coupled YIG/Py magnetization dynamics are excited by directly applying microwave current through the conductive Py layer. Only the Py layer contributes to the spin rectification signal because the YIG is insulating, enabling clear phase-resolved detection of the Py component of the YIG-Py hybrid modes. We measure a constant phase for the Py-dominated hybrid modes and a $\pi$ phase offset across the avoided crossing for the YIG-dominated modes. From the slope of the $\pi$ phase shift we can determine the interlayer coupling strength, agreeing with the measurement from the avoided crossing. We have also characterized a nonlocally excited YIG/Py sample, in which a phase offset compared with the single bilayer device suggests the existence of a large perpendicular Oersted field driving the dynamics. Our results open an avenue of building up, reading out and designing circuits of on-chip magnonic hybrid devices for the application of coherent magnonic information processing.

\begin{figure*}[htb]
 \centering
 \includegraphics[width=5.5 in]{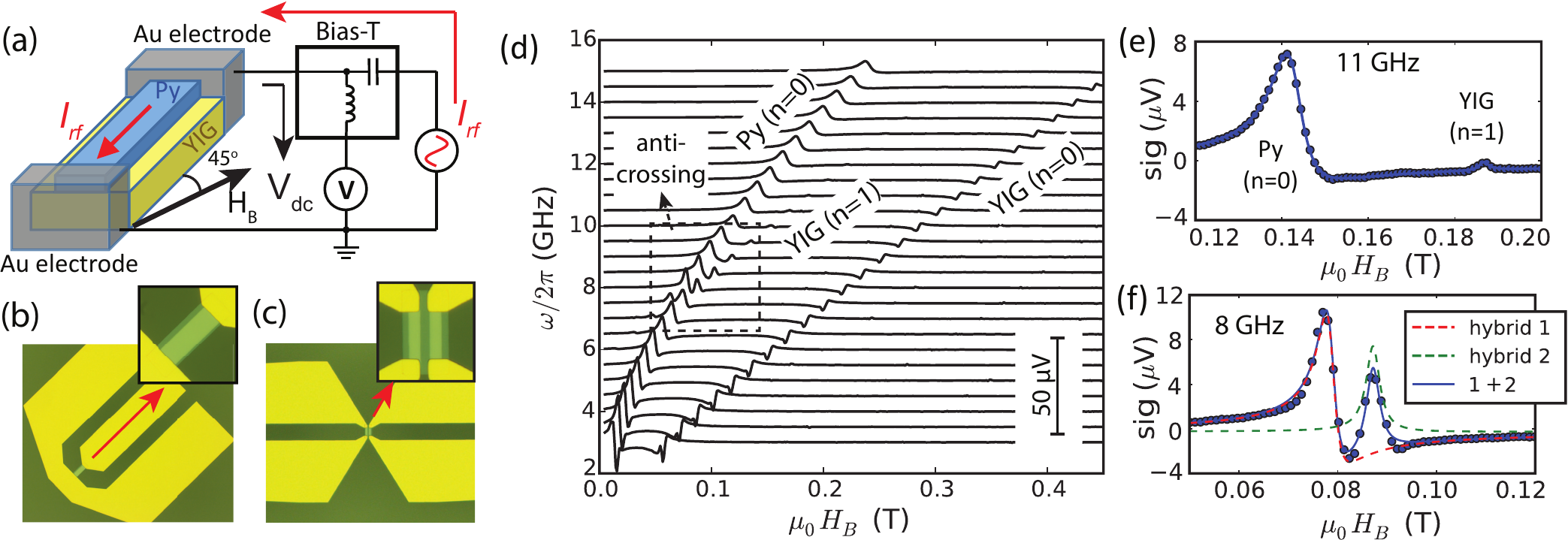}
 \caption{(a) Illustration of electrical excitation and detection of YIG/Py bilayer devices. The in-plane external biasing field is kept as 45$^\circ$ from the $I_{rf}$ direction along the Py devices. (b-c) Optical microscope images of the device and Au coplanar waveguide antenna for (b) single devices and (c) nonlocally excited device. \textcolor{black}{(d-f) Spin rectification signals for the YIG(70~nm)/Py(9~nm) single device, with the mode anti-crossing between YIG ($n=1$) and Py ($n=0$) modes marked by the dashed box. (e) Zoom-in lineshape of (d) at 11 GHz where the extrapolated YIG ($n=1$) and Py ($n=0$) peaks are well separated. (f) Lineshape at 8 GHz where the YIG ($n=1$) and Py ($n=0$) modes are degenerate in field. The red and green dashed curves in (f) denote the fit of the two YIG-Py hybrid modes.}}
 \label{fig1}
\end{figure*}

YIG thin films (50~nm, 70~nm and 85~nm) were sputtered on Gd$_3$Ga$_5$O$_{12}$ (111) substrate with lithographically defined device patterns, followed by liftoff and annealing in air at 850 $^\circ$C for 3 h \cite{LiNanoscale2016,LiPRL2020_YIGPy}. Then a second-layer Py device (8~nm or 9~nm) was defined on the YIG device with lithography and sputtering, with 1 min ion milling of YIG surface in vacuum right before deposition. Lastly a 200~nm thick Au coplanar waveguide (CPW) was fabricated which was in contact with the Py device for electrical excitations and measurements. Fig.~\ref{fig1}(a) shows the schematic of the spin rectification measurement. The top-view optical microscope images of the devices are shown in Fig.~\ref{fig1}(b) for the single devices and (c) for the nonlocally excited devices. The dimensions of the Py devices are 10~$\mu$m $\times$ 40~$\mu$m in Fig.~\ref{fig1}(b) and 6~$\mu$m $\times$ 20~$\mu$m in Fig.~\ref{fig1}(c). The two Py devices in Figs. \ref{fig1}(c) are separated by 2~$\mu$m, one for applying nonlocal excitation signals and the other for the spin rectification measurements. Throughout the measurements, the external biasing field is applied in the sample plane and tilted 45$^\circ$ away from the microwave current direction, which is the commonly used configuration in spin rectification measurements for maximizing the output signals \cite{SklenarPRB2015}.

\begin{figure}[htb]
 \centering
 \includegraphics[width=3.0 in]{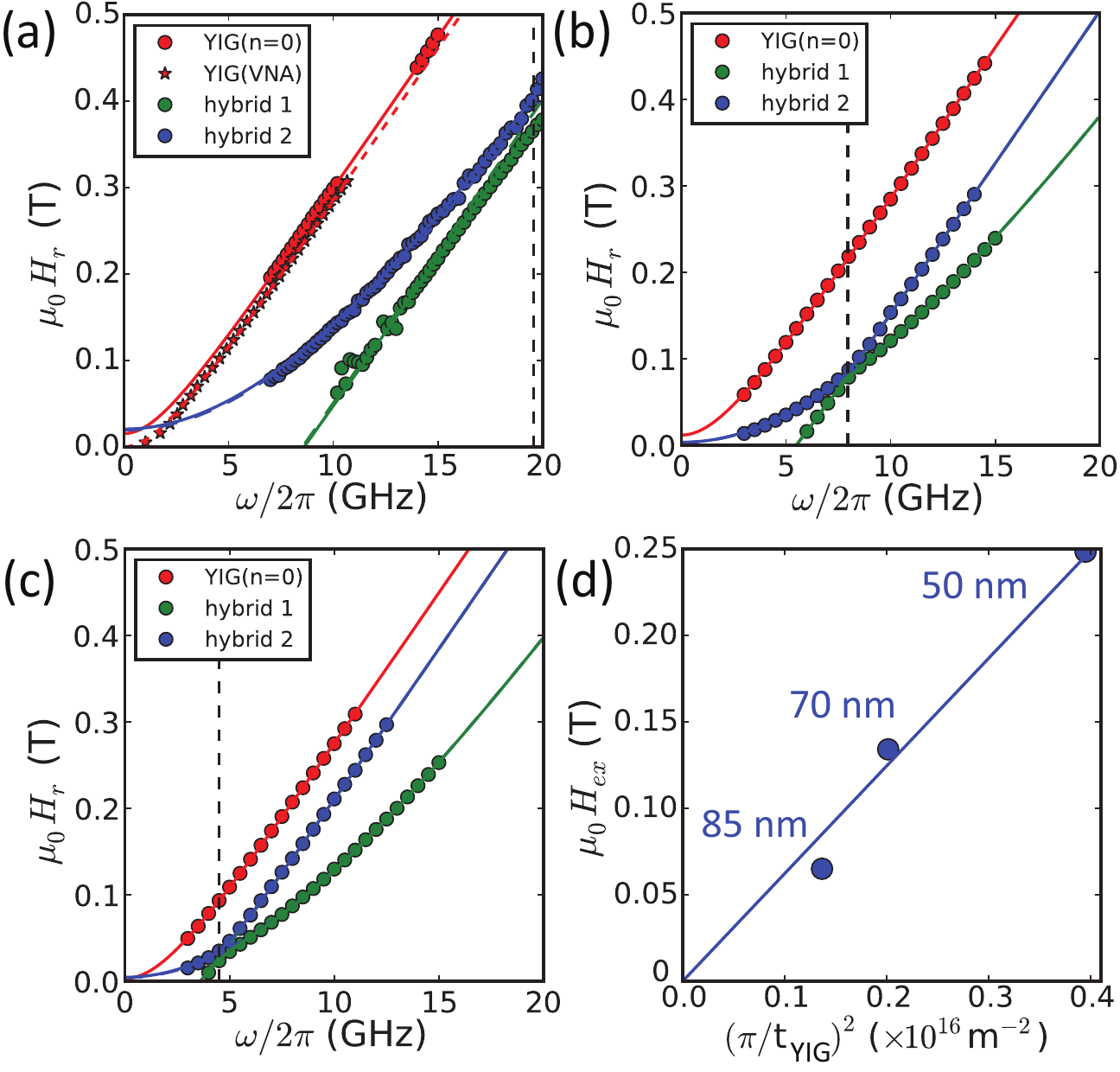}
 \caption{(a-c) Extracted resonance peak positions of (a) YIG(50~nm)/Py(8~nm), (b) YIG(70~nm)/Py(9~nm) and (c) YIG(85~nm)/Py(9~nm) single devices. The mode degeneracy between the Py uniform mode and YIG ($n=1$) mode happens at $\omega_c/2\pi=19.5$~GHz for (a), 7.9~GHz for (b) and 4.7~GHz for (c), denoted by vertical dashed lines. (d) Exchange field for different $t_\text{YIG}$.}
 \label{fig2}
\end{figure}

Fig.~\ref{fig1}(d) shows the field-swept spin rectification signals of the YIG(70~nm)/Py(9~nm) single device at different frequencies. We observe both the nominal Py and YIG uniform mode resonances, as reported previously \cite{HydePRB2014,YangPRB2020}, even though the signals come from only the Py layer. For the Py uniform mode, the excitation is mainly due to a finite Oersted field projection to the dynamic mode, which has also been observed in a single CoFe layer in our prior work \cite{LiPRL2019_CoFe}. For the YIG excitation, the interfacial exchange coupling creates coupled modes with finite amplitude on the Py, leading to a modulation of the Py resistance even for mode that is nominally the YIG uniform mode. In addition, the YIG ($n=1$) PSSW mode is also excited when it intersects with the Py uniform mode, forming an avoided crossing between the two modes at $\omega_c/2\pi=7.9$ GHz \textcolor{black}{(Fig. \ref{fig1}f)}. \textcolor{black}{The separation of the two hybrid peaks is 8.5 mT, which is larger than the extrapolated individual linewidths of the Py ($n=0$) and YIG($n=1$) modes ($\mu_0 \Delta H_\text{Py}=5.5$ mT, $\mu_0 \Delta H_\text{YIG}=3.0$ mT)}. Far away from the avoided crossing, the excitations of the YIG ($n=1$) mode are almost unnoticeable \textcolor{black}{(Fig. \ref{fig1}e)}, which is due to the weak coupling of the uniform Oersted field to the odd PSSW modes. Thus the drive of the YIG ($n=1$) mode is dominated by excitation of the admixture of the Py mode due to the interfacial exchange \cite{KlinglerPRL2018,ChenPRL2018,QinSREP2018,LiPRL2020_YIGPy}.

To analyze the spin rectification signals, the measured lineshape for each peak can be fitted to the following function:
\begin{equation}\label{eq00}
V_\text{dc} = \operatorname{Im}\left[{V_0 e^{i\phi}\Delta H \over (H_B-H_r)-i\Delta H}\right]
\end{equation}
where $H_B$ is the biasing field, $H_r$ is the resonance field as a function of frequency, $\Delta H$ is the half-width-half-maximum linewidth, $V_0$ is the peak amplitude, and $\phi$ represents the mixing of the symmetric and antisymmetric Lorentzian lineshapes and reflects the phase evolution of the Py component in the YIG-Py hybrid dynamics. The operator $\operatorname{Im}[]$ takes the imaginary part. This technique has been used to probe the dampinglike and fieldlike torque components \cite{ChenAPL2008,FangNNano2011,LiuPRL2011,BaiPRL2013,MellnikNature2014,ZhangWeiPRL2014,SanchezPRL2014,SklenarPRB2015,NanPRB2015,JungfleischPRL2016} as well as in recent optical rectification experiments \cite{HisatomiPRB2016,YoonPRB2016,CapuaPRB2017,LiPRApplied2019,LiIEEETransMagn2019,WeiJinwuPRApplied2020}. Moreover, the single source of spin rectification signal from Py allows convenient theoretical analysis for studying the phase evolution of the hybrid dynamics, as will be shown below.

Figs. \ref{fig2}(a-c) show the extracted $H_r$ as a function of frequency $\omega$ for $t_\text{YIG}=50$~nm, 70~nm and 85~nm, respectively. The two hybrid modes, marked as blue and green circles, are formed between Py uniform and YIG ($n=1$) modes. With different $t_\text{YIG}$, the mode intersection happens at different frequencies due to the effective exchange field $\mu_0H_{ex}=(2A_{ex}/M_s)k^2$ with $k=\pi/t_\text{YIG}$, which shifts the YIG ($n=1$) mode towards higher frequencies. Fig.~\ref{fig2}(d) plots the extracted $\mu_0H_{ex}$ as a function of $(\pi/t_\text{YIG})^2$; good linear dependence confirms the role of the exchange field. By fitting the data to the Kittel equation plus the exchange field, we obtain similar values of magnetization in all films as $\mu_0M_\text{Py}=0.81$ T, $\mu_0M_\text{YIG}=0.19$ T. From the linear fits in Fig.~\ref{fig2}(d) we obtain $A_{ex}=4.7$ pJ/m for the YIG film.

The mode anticrossing behaviors in Figs. \ref{fig2}(a-c) can be fitted to the equation developed by the two coupled magnon resonances \cite{LiPRL2020_YIGPy}:
\begin{equation}\label{eq01}
\mu_0H_\pm = \mu_0{H_\text{YIG}+H_\text{Py} \over 2} \pm \sqrt{ \left(\mu_0{H_\text{YIG}-H_\text{Py} \over 2}\right)^2 + g_H^2}
\end{equation}
where $\mu_0H_\textrm{YIG}=\sqrt{\mu_0^2M_\textrm{YIG}^2/4+\omega^2/\gamma^2}-\mu_0M_\textrm{YIG}/2 +\mu_0H_{ex}$ and $\mu_0H_\textrm{Py}=\sqrt{\mu_0^2M_\textrm{Py}^2/4+\omega^2/\gamma^2}-\mu_0M_\textrm{Py}/2$ are the solutions of the Kittel equation for the YIG ($n=1$) and Py modes, $\gamma/2\pi= (g_{eff}/2)\times 27.99$~GHz/T and $g_H$ is the interfacial exchange coupling strength in the magnetic field domain. In our previous work \cite{LiPRL2020_YIGPy}, we derived that $g_H = f(\omega)\sqrt{J/M_\text{Py}t_\text{Py}\cdot J/M_\text{YIG}t_\text{YIG}}$ where $J$ is the interfacial exchange coupling strength. \textcolor{black}{The factor $f(\omega)\approx 0.9$ accounts for the nonlinearity due to the demagnetizing field}. Data fittings to Eq. (\ref{eq01}) yield an averaged $g_H=8.7$~mT and $J=0.066$~mJ/m$^2$; the latter is consistent with the reported value of 0.06~mJ/m$^2$ for continuous thin films \cite{LiPRL2020_YIGPy}. We also double-check the value of $J$ by measuring the inductive ferromagnetic resonance of a 200~$\mu$m $\times$ 40~$\mu$m YIG stripe from the same fabrication as the YIG(50~nm)/Py(8~nm) device, with the peak dispersion shown as stars in Fig.~\ref{fig2}(a). From the Kittel fitting, we obtain a constant resonance field offset of $\mu_0H_k=13.8$~mT between the YIG stripe and the YIG/Py device. From this static offset, we extract $J=\mu_0H_k M_\text{YIG}t_\text{YIG}=0.110$ mJ/m$^2$ \cite{LiPRL2020_YIGPy}, in good agreement with the value of 0.112 mJ/m$^2$ obtained above from the avoided crossing for $t_\text{YIG}=50$~nm. The YIG/Py device shows a higher resonance field than YIG, confirming the antiferromagnetic exchange coupling between YIG and Py.

\begin{figure*}[htb]
 \centering
 \includegraphics[width=6.0 in]{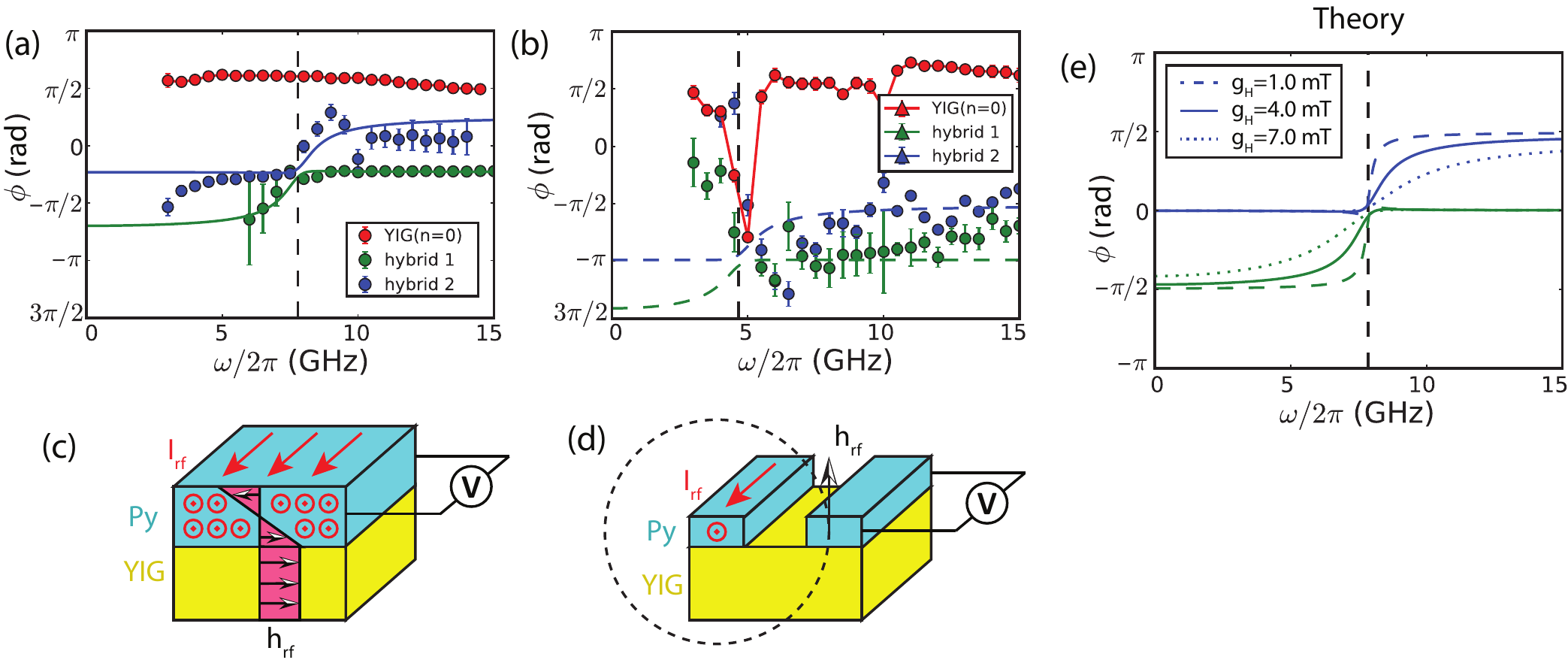}
 \caption{Phase evolution of the spin rectification signals for (a) YIG(70~nm)/Py(9~nm) single device and (b) YIG(85~nm)/Py(9~nm) nonlocally excited device, with their microwave current flow and field distribution illustrated in (c) and (d), respectively. The blue and green curves show the theoretical prediction from Eq. (\ref{eq04}) with (b) $g_H= 4.0$ mT for (a) and 5.3 mT for (b). The error bars indicate single standard deviation uncertainties that arise primarily from the fitting of the resonances. (e) Theoretical plots of phase evolution from Eq. (\ref{eq04}) using the $H_r$ in (a) and $\phi_\text{Py}=0$ for different $g_H$.}
 \label{fig3}
\end{figure*}

Next, we show the evolution of $\phi$ for the hybrid modes, which are the main results of this work. Fig.~\ref{fig3}(a) shows the extracted phases for the three modes in the YIG(70~nm)/Py(9~nm) single device, with the color corresponding to the resonance field plot in Fig.~\ref{fig2}(b). The two hybrid modes exhibit a clear phase crossing where their resonance fields intersect at $\omega_c/2\pi=7.9$~GHz (vertical dashed line). For the Py-dominated hybrid modes which are represented by the blue circles lower than $\omega_c$ and the green circles higher than $\omega_c$, the phase stays at a constant level ($\phi_\text{Py}=-0.23$ $\pi$). This is expected in spin rectification measurements, where a consistent phase relation between the Py dynamics and the microwave current is maintained in a broad frequency domain. For ideal fieldlike excitations as illustrated in Fig.~\ref{fig3}(c) in the single device, we expect $\phi_\text{Py}=-\pi/2$. Experimentally, the deviation of $\phi_\text{Py}$ may be due to the self spin torque providing a finite dampinglike drive component \cite{WangNNano2019}. Alternatively, the phase offset may be also a reflection of the inhomogeneous mode profile of Py in the presence of the YIG/Py interfacial exchange boundary as well as the nonuniform current distribution across the thickness of Py.

The phase of the YIG-dominated hybrid modes, on the other hand, evolve from below $\phi_\text{Py}$ to above $\phi_\text{Py}$ with an increment of nearly $\pi$. As a rough explanation, by passing through the avoided crossing, the frequency of the YIG-dominated mode evolves from below the Py resonance frequency to above it. This leads to a phase shift of $\pi$ for the Py susceptibility. Because the YIG dynamics is driven by the interfacial exchange from the Py excitation, a phase shift of $\pi$ is also expected in the YIG-dominated mode. Furthermore, due to the strong magnon-magnon coupling, the $\pi$ phase shift does not take a sharp transition at $\omega_c$, but takes a gradual transition with the transition bandwidth determined by the coupling strength $g_H$.

To quantitatively understand the phase evolution of the hybrid mode, we follow the susceptibility tensor which has been derived in our prior work {see Eq. (S-4) in the Supplemental Materials of Ref. \cite{LiPRL2020_YIGPy}}. In the limit of weak damping and ignoring the precession ellipticity, the dynamics of the Py uniform and YIG ($n=1$) modes can be expressed as:
\begin{subequations}\label{eq02}
\begin{align}
  \tilde{m}_\textrm{Py}&= {\tilde{h}^x_\textrm{Py} \over {H_B-H_\textrm{Py}-i\Delta H_\textrm{Py}-{g_H^2 \over {H_B-H_\textrm{YIG}-i\Delta H_\textrm{YIG}}}}}\\
  \tilde{m}_\textrm{YIG}&= {g_H\tilde{m}_\textrm{Py} \over {H_B-H_\textrm{YIG}-i\Delta H_\textrm{YIG}}}
\end{align}
\end{subequations}
where $\tilde{m}_\text{Py}$ and $\tilde{m}_\text{YIG}$ denote the unitless transverse components for Py and YIG, $\Delta H_\textrm{Py}$ and $\Delta H_\textrm{YIG}$ denote their linewidths. For the Py layer, the effective field $\tilde{h}^x_\text{Py}$ is exerted from the microwave current flowing through. For the YIG layer, the effective field $g_H\tilde{m}_\textrm{Py}$ is provided by the interfacial exchange when the Py magnetization precesses. Note that because YIG is an insulator, the spin rectification signal is only contributed by $\tilde{m}_\textrm{Py}$, which significantly simplifies the theoretical analysis. Eq. (\ref{eq02}a) can be rewritten as:
\begin{equation}\label{eq03}
  \tilde{m}_\text{Py}= {\tilde{h}^x_\text{Py}(H_B-H_\textrm{YIG}-i\Delta H_\textrm{YIG}) \over (H_B-H_+-i\Delta H_+)(H_B-H_--i\Delta H_-) }
\end{equation}
where the values of $H_\pm$ are defined in Eq. (\ref{eq01}) and $\Delta H_\pm$ are the linewidths for the two hybrid modes. Compared with Eq. (\ref{eq00}), the phase for the $H_\pm$ resonance can be finally expressed as:
\begin{equation}\label{eq04}
\phi_\pm = \phi_\text{Py} + \tan^{-1}\left( -\Delta H_\text{YIG} \over H_\pm-H_\text{YIG} \right) - \tan^{-1}\left( -\Delta H_\mp \over H_\pm-H_\mp \right)
\end{equation}
In Eq. (\ref{eq04}) the first term comes from a finite phase offset between $\tilde{h}^x_\text{Py}$ and the microwave current, the second term comes from the numerator and provides the $\pi$ phase shift, and the last term is usually close to zero in the strong coupling regime as the linewidth is much smaller than the resonance detuning. The calculation results of Eq. (\ref{eq04}) are plotted in Fig.~\ref{fig3}(a), which nicely reproduce the experimental data and the positive increment of phase for the YIG-dominated hybrid mode. We also plot the calculated phase evolution for different values of $g_H$ in Fig.~\ref{fig3}(e). For small $g_H$, the YIG-dominated mode shows a rapid phase shift near the mode crossing frequency. As $g_H$ increases, the phase transition regime broadens because $g_H$ defines how quickly the hybrid mode evolves to uncoupled individual modes.

The phase-resolved spin rectification measurement of the hybrid modes are also repeated on a nonlocally excited device. With the excitation and detection schematics shown in Fig.~\ref{fig3}(d), the microwave current flows through a nonlocal Py electrode, which provides an Oersted field that is perpendicular to the Py device being measured. For the detection, due to the inductive coupling between the two adjacent Py devices, a finite microwave current flows through the second Py device which leads to a measurable spin rectification voltage when the Py magnetization dynamics is excited. Fig.~\ref{fig3}(b) shows the measured $\phi$ for the three modes. Above $\omega_c/2\pi=4.7$~GHz, the YIG-dominated mode exhibits a phase advance close to $\pi/2$ compared with the Py-dominated mode, which agrees with the theoretical prediction. For the Py-dominated mode, the extracted value of $\phi_\text{Py}=-0.99$ $\pi$ also agrees with theoretical prediction of $\phi_\text{Py}=-\pi$ due to the additional $-\pi/2$ phase offset from the perpendicular Oersted field from the nonlocal antenna. Below 4.7 GHz, the anomalous phase drift is accompanied with the linewidth drift and is due to the weak signals. Thus we consider this low-frequency phase drift as to be an artifact due to weak signals rather than a significant effect. Note that the nonlocal excitation schematic should eliminate the spurious phase offset due to the complex excitation profile, because the out-of-plane Oersted field is rather uniform.

The YIG uniform modes exhibit a consistent phase of $\phi_\text{Py}=\pi/2$ in both Figs.~\ref{fig3}(a) and (b). Note that we still use $\phi_\text{Py}$ to represent the phase because the spin rectification signals come from the motion of the Py layer induced by the resonance of YIG via the interfacial exchange \cite{ChibaPRApplied2014,SklenarPRB2015}. The value of $\phi_\text{Py}$ suggests a dominating in-plane Oersted field on the YIG layer from the microwave current flowing through the adjacent Py layer. For the YIG(70~nm)/Py(9~nm) single device, the only Py layer acts as an antenna which is highly efficient in exciting the YIG uniform mode [Fig.~\ref{fig3}(c)]. For the YIG(85~nm)/Py(9~nm) nonlocally excited device, the unchanged $\phi_\text{Py}=\pi/2$ shows that the perpendicular field from the nonlocal Py antenna is still insignificant compared with the induced microwave current in the Py device being electrically measured, with the latter much more efficient in producing an in-plane Oersted field on the YIG layer underneath. Note that the sign change of $\phi_\text{Py}$ from the Py-dominated uniform mode is caused by the negative value of $g_H$ from antiferromagnetic coupling, adding an additional $\pi$ phase to the YIG uniform mode. A similar observation has also been reported in Ref. \cite{YangPRB2020}.

In conclusion, we have demonstrated phase-resolved electrical measurements of YIG/Py bilayer devices with strong magnon-magnon coupling. The micron-wide and nanometer-thick devices serve as an on-chip miniaturized two-cavity hybrid system, where the two microwave cavities are composed of two exchange-coupled thin layers of magnon resonators. Furthermore, the unique coupling mechanism and the confined magnon resonance allow versatile geometric configuration, such as the nonlocal device, as well as convenient electrical excitation and detection. In the recent rapid development of cavity spintronics and magnon hybrid systems \cite{BhoiSSP2019,LachanceQuirionAPEx2019,KusminskiyarXiv2019,WangJAP2020,LiJAP2020}, lots of emerging physics and device engineering including exceptional points \cite{ZhangDengkeNComm2017,ZhangPRL2019}, level attraction \cite{HarderPRL2018,BhoiPRB2019,BoventerPRResearch2020} and nonreciprocity \cite{WangPRL2019,ZhangPRApplied2020} have utilized coherent interaction of different microwave ingredients. Our results provide a platform for implementing and realizing these findings in geometrically confined, thin-film based dynamic systems and for studying the driving and coupling interactions, which are critical for applications in coherent information processing.

Work at Argonne on sample preparation and characterization was supported by the U.S. Department of Energy, Office of Science, Basic Energy Sciences, Materials Sciences and Engineering Division, while work at Argonne and National Institute of Standards and Technology (NIST) on data analysis and theoretical modeling was supported as part of Quantum Materials for Energy Efficient Neuromorphic Computing, an Energy Frontier Research Center funded by the U.S. DOE, Office of Science, Basic Energy Sciences (BES) under Award \#DE-SC0019273. Use of the Center for Nanoscale Materials, an Office of Science user facility, was supported by the U.S. Department of Energy, Office of Science, Office of Basic Energy Sciences, under Contract No. DE-AC02-06CH11357. W. Z. acknowledges support from AFOSR under grant no. FA9550-19-1-0254.

\section{DATA AVAILABILITY}
The data that support the findings of this study are available from the corresponding author upon reasonable request.


\begin{thebibliography}{71}%
\makeatletter
\providecommand \@ifxundefined [1]{%
 \@ifx{#1\undefined}
}%
\providecommand \@ifnum [1]{%
 \ifnum #1\expandafter \@firstoftwo
 \else \expandafter \@secondoftwo
 \fi
}%
\providecommand \@ifx [1]{%
 \ifx #1\expandafter \@firstoftwo
 \else \expandafter \@secondoftwo
 \fi
}%
\providecommand \natexlab [1]{#1}%
\providecommand \enquote  [1]{``#1''}%
\providecommand \bibnamefont  [1]{#1}%
\providecommand \bibfnamefont [1]{#1}%
\providecommand \citenamefont [1]{#1}%
\providecommand \href@noop [0]{\@secondoftwo}%
\providecommand \href [0]{\begingroup \@sanitize@url \@href}%
\providecommand \@href[1]{\@@startlink{#1}\@@href}%
\providecommand \@@href[1]{\endgroup#1\@@endlink}%
\providecommand \@sanitize@url [0]{\catcode `\\12\catcode `\$12\catcode
  `\&12\catcode `\#12\catcode `\^12\catcode `\_12\catcode `\%12\relax}%
\providecommand \@@startlink[1]{}%
\providecommand \@@endlink[0]{}%
\providecommand \url  [0]{\begingroup\@sanitize@url \@url }%
\providecommand \@url [1]{\endgroup\@href {#1}{\urlprefix }}%
\providecommand \urlprefix  [0]{URL }%
\providecommand \Eprint [0]{\href }%
\providecommand \doibase [0]{http://dx.doi.org/}%
\providecommand \selectlanguage [0]{\@gobble}%
\providecommand \bibinfo  [0]{\@secondoftwo}%
\providecommand \bibfield  [0]{\@secondoftwo}%
\providecommand \translation [1]{[#1]}%
\providecommand \BibitemOpen [0]{}%
\providecommand \bibitemStop [0]{}%
\providecommand \bibitemNoStop [0]{.\EOS\space}%
\providecommand \EOS [0]{\spacefactor3000\relax}%
\providecommand \BibitemShut  [1]{\csname bibitem#1\endcsname}%
\let\auto@bib@innerbib\@empty
\bibitem [{\citenamefont {Huebl}\ \emph {et~al.}(2013)\citenamefont {Huebl},
  \citenamefont {Zollitsch}, \citenamefont {Lotze}, \citenamefont {Hocke},
  \citenamefont {Greifenstein}, \citenamefont {Marx}, \citenamefont {Gross},\
  and\ \citenamefont {Goennenwein}}]{HueblPRL2013}%
  \BibitemOpen
  \bibfield  {author} {\bibinfo {author} {\bibfnamefont {H.}~\bibnamefont
  {Huebl}}, \bibinfo {author} {\bibfnamefont {C.~W.}\ \bibnamefont
  {Zollitsch}}, \bibinfo {author} {\bibfnamefont {J.}~\bibnamefont {Lotze}},
  \bibinfo {author} {\bibfnamefont {F.}~\bibnamefont {Hocke}}, \bibinfo
  {author} {\bibfnamefont {M.}~\bibnamefont {Greifenstein}}, \bibinfo {author}
  {\bibfnamefont {A.}~\bibnamefont {Marx}}, \bibinfo {author} {\bibfnamefont
  {R.}~\bibnamefont {Gross}}, \ and\ \bibinfo {author} {\bibfnamefont
  {S.~T.~B.}\ \bibnamefont {Goennenwein}},\ }\href {\doibase
  10.1103/PhysRevLett.111.127003} {\bibfield  {journal} {\bibinfo  {journal}
  {Phys. Rev. Lett.}\ }\textbf {\bibinfo {volume} {111}},\ \bibinfo {pages}
  {127003} (\bibinfo {year} {2013})}\BibitemShut {NoStop}%
\bibitem [{\citenamefont {Tabuchi}\ \emph {et~al.}(2014)\citenamefont
  {Tabuchi}, \citenamefont {Ishino}, \citenamefont {Ishikawa}, \citenamefont
  {Yamazaki}, \citenamefont {Usami},\ and\ \citenamefont
  {Nakamura}}]{TabuchiPRL2014}%
  \BibitemOpen
  \bibfield  {author} {\bibinfo {author} {\bibfnamefont {Y.}~\bibnamefont
  {Tabuchi}}, \bibinfo {author} {\bibfnamefont {S.}~\bibnamefont {Ishino}},
  \bibinfo {author} {\bibfnamefont {T.}~\bibnamefont {Ishikawa}}, \bibinfo
  {author} {\bibfnamefont {R.}~\bibnamefont {Yamazaki}}, \bibinfo {author}
  {\bibfnamefont {K.}~\bibnamefont {Usami}}, \ and\ \bibinfo {author}
  {\bibfnamefont {Y.}~\bibnamefont {Nakamura}},\ }\href {\doibase
  10.1103/PhysRevLett.113.083603} {\bibfield  {journal} {\bibinfo  {journal}
  {Phys. Rev. Lett.}\ }\textbf {\bibinfo {volume} {113}},\ \bibinfo {pages}
  {083603} (\bibinfo {year} {2014})}\BibitemShut {NoStop}%
\bibitem [{\citenamefont {Zhang}\ \emph
  {et~al.}(2014{\natexlab{a}})\citenamefont {Zhang}, \citenamefont {Zou},
  \citenamefont {Jiang},\ and\ \citenamefont {Tang}}]{ZhangPRL2014}%
  \BibitemOpen
  \bibfield  {author} {\bibinfo {author} {\bibfnamefont {X.}~\bibnamefont
  {Zhang}}, \bibinfo {author} {\bibfnamefont {C.-L.}\ \bibnamefont {Zou}},
  \bibinfo {author} {\bibfnamefont {L.}~\bibnamefont {Jiang}}, \ and\ \bibinfo
  {author} {\bibfnamefont {H.~X.}\ \bibnamefont {Tang}},\ }\href {\doibase
  10.1103/PhysRevLett.113.156401} {\bibfield  {journal} {\bibinfo  {journal}
  {Phys. Rev. Lett.}\ }\textbf {\bibinfo {volume} {113}},\ \bibinfo {pages}
  {156401} (\bibinfo {year} {2014}{\natexlab{a}})}\BibitemShut {NoStop}%
\bibitem [{\citenamefont {Goryachev}\ \emph {et~al.}(2014)\citenamefont
  {Goryachev}, \citenamefont {Farr}, \citenamefont {Creedon}, \citenamefont
  {Fan}, \citenamefont {Kostylev},\ and\ \citenamefont
  {Tobar}}]{GoryachevPRApplied2014}%
  \BibitemOpen
  \bibfield  {author} {\bibinfo {author} {\bibfnamefont {M.}~\bibnamefont
  {Goryachev}}, \bibinfo {author} {\bibfnamefont {W.~G.}\ \bibnamefont {Farr}},
  \bibinfo {author} {\bibfnamefont {D.~L.}\ \bibnamefont {Creedon}}, \bibinfo
  {author} {\bibfnamefont {Y.}~\bibnamefont {Fan}}, \bibinfo {author}
  {\bibfnamefont {M.}~\bibnamefont {Kostylev}}, \ and\ \bibinfo {author}
  {\bibfnamefont {M.~E.}\ \bibnamefont {Tobar}},\ }\href {\doibase
  10.1103/PhysRevApplied.2.054002} {\bibfield  {journal} {\bibinfo  {journal}
  {Phys. Rev. Applied}\ }\textbf {\bibinfo {volume} {2}},\ \bibinfo {pages}
  {054002} (\bibinfo {year} {2014})}\BibitemShut {NoStop}%
\bibitem [{\citenamefont {Bhoi}\ \emph {et~al.}(2014)\citenamefont {Bhoi},
  \citenamefont {Cliff}, \citenamefont {Maksymov}, \citenamefont {Kostylev},
  \citenamefont {Aiyar}, \citenamefont {Venkataramani}, \citenamefont
  {Prasad},\ and\ \citenamefont {Stamps}}]{BhoiJAP2014}%
  \BibitemOpen
  \bibfield  {author} {\bibinfo {author} {\bibfnamefont {B.}~\bibnamefont
  {Bhoi}}, \bibinfo {author} {\bibfnamefont {T.}~\bibnamefont {Cliff}},
  \bibinfo {author} {\bibfnamefont {I.~S.}\ \bibnamefont {Maksymov}}, \bibinfo
  {author} {\bibfnamefont {M.}~\bibnamefont {Kostylev}}, \bibinfo {author}
  {\bibfnamefont {R.}~\bibnamefont {Aiyar}}, \bibinfo {author} {\bibfnamefont
  {N.}~\bibnamefont {Venkataramani}}, \bibinfo {author} {\bibfnamefont
  {S.}~\bibnamefont {Prasad}}, \ and\ \bibinfo {author} {\bibfnamefont {R.~L.}\
  \bibnamefont {Stamps}},\ }\href {\doibase 10.1063/1.4904857} {\bibfield
  {journal} {\bibinfo  {journal} {J. Appl. Phys.}\ }\textbf {\bibinfo {volume}
  {116}},\ \bibinfo {pages} {243906} (\bibinfo {year} {2014})}\BibitemShut
  {NoStop}%
\bibitem [{\citenamefont {Tabuchi}\ \emph {et~al.}(2015)\citenamefont
  {Tabuchi}, \citenamefont {Ishino}, \citenamefont {Noguchi}, \citenamefont
  {Ishikawa}, \citenamefont {Yamazaki}, \citenamefont {Usami},\ and\
  \citenamefont {Nakamura}}]{TabuchiScience2015}%
  \BibitemOpen
  \bibfield  {author} {\bibinfo {author} {\bibfnamefont {Y.}~\bibnamefont
  {Tabuchi}}, \bibinfo {author} {\bibfnamefont {S.}~\bibnamefont {Ishino}},
  \bibinfo {author} {\bibfnamefont {A.}~\bibnamefont {Noguchi}}, \bibinfo
  {author} {\bibfnamefont {T.}~\bibnamefont {Ishikawa}}, \bibinfo {author}
  {\bibfnamefont {R.}~\bibnamefont {Yamazaki}}, \bibinfo {author}
  {\bibfnamefont {K.}~\bibnamefont {Usami}}, \ and\ \bibinfo {author}
  {\bibfnamefont {Y.}~\bibnamefont {Nakamura}},\ }\href {\doibase
  10.1126/science.aaa3693} {\bibfield  {journal} {\bibinfo  {journal}
  {Science}\ }\textbf {\bibinfo {volume} {349}},\ \bibinfo {pages} {405}
  (\bibinfo {year} {2015})}\BibitemShut {NoStop}%
\bibitem [{\citenamefont {Zhang}\ \emph {et~al.}(2015)\citenamefont {Zhang},
  \citenamefont {Zou}, \citenamefont {Zhu}, \citenamefont {Marquardt},
  \citenamefont {Jiang},\ and\ \citenamefont {Tang}}]{ZhangNComm2015}%
  \BibitemOpen
  \bibfield  {author} {\bibinfo {author} {\bibfnamefont {X.}~\bibnamefont
  {Zhang}}, \bibinfo {author} {\bibfnamefont {C.-L.}\ \bibnamefont {Zou}},
  \bibinfo {author} {\bibfnamefont {N.}~\bibnamefont {Zhu}}, \bibinfo {author}
  {\bibfnamefont {F.}~\bibnamefont {Marquardt}}, \bibinfo {author}
  {\bibfnamefont {L.}~\bibnamefont {Jiang}}, \ and\ \bibinfo {author}
  {\bibfnamefont {H.~X.}\ \bibnamefont {Tang}},\ }\href@noop {} {\bibfield
  {journal} {\bibinfo  {journal} {Nature Communi.}\ }\textbf {\bibinfo {volume}
  {6}},\ \bibinfo {pages} {8914} (\bibinfo {year} {2015})}\BibitemShut
  {NoStop}%
\bibitem [{\citenamefont {Bai}\ \emph {et~al.}(2015)\citenamefont {Bai},
  \citenamefont {Harder}, \citenamefont {Chen}, \citenamefont {Fan},
  \citenamefont {Xiao},\ and\ \citenamefont {Hu}}]{BaiPRL2015}%
  \BibitemOpen
  \bibfield  {author} {\bibinfo {author} {\bibfnamefont {L.}~\bibnamefont
  {Bai}}, \bibinfo {author} {\bibfnamefont {M.}~\bibnamefont {Harder}},
  \bibinfo {author} {\bibfnamefont {Y.~P.}\ \bibnamefont {Chen}}, \bibinfo
  {author} {\bibfnamefont {X.}~\bibnamefont {Fan}}, \bibinfo {author}
  {\bibfnamefont {J.~Q.}\ \bibnamefont {Xiao}}, \ and\ \bibinfo {author}
  {\bibfnamefont {C.-M.}\ \bibnamefont {Hu}},\ }\href {\doibase
  10.1103/PhysRevLett.114.227201} {\bibfield  {journal} {\bibinfo  {journal}
  {Phys. Rev. Lett.}\ }\textbf {\bibinfo {volume} {114}},\ \bibinfo {pages}
  {227201} (\bibinfo {year} {2015})}\BibitemShut {NoStop}%
\bibitem [{\citenamefont {Lachance-Quirion}\ \emph {et~al.}(2017)\citenamefont
  {Lachance-Quirion}, \citenamefont {Tabuchi}, \citenamefont {Ishino},
  \citenamefont {Noguchi}, \citenamefont {Ishikawa}, \citenamefont {Yamazaki},\
  and\ \citenamefont {Nakamura}}]{LachanceScienceAdvan2017}%
  \BibitemOpen
  \bibfield  {author} {\bibinfo {author} {\bibfnamefont {D.}~\bibnamefont
  {Lachance-Quirion}}, \bibinfo {author} {\bibfnamefont {Y.}~\bibnamefont
  {Tabuchi}}, \bibinfo {author} {\bibfnamefont {S.}~\bibnamefont {Ishino}},
  \bibinfo {author} {\bibfnamefont {A.}~\bibnamefont {Noguchi}}, \bibinfo
  {author} {\bibfnamefont {T.}~\bibnamefont {Ishikawa}}, \bibinfo {author}
  {\bibfnamefont {R.}~\bibnamefont {Yamazaki}}, \ and\ \bibinfo {author}
  {\bibfnamefont {Y.}~\bibnamefont {Nakamura}},\ }\href {\doibase
  10.1126/sciadv.1603150} {\bibfield  {journal} {\bibinfo  {journal} {Science
  Advances}\ }\textbf {\bibinfo {volume} {3}} (\bibinfo {year} {2017}),\
  10.1126/sciadv.1603150}\BibitemShut {NoStop}%
\bibitem [{\citenamefont {Bai}\ \emph {et~al.}(2017)\citenamefont {Bai},
  \citenamefont {Harder}, \citenamefont {Hyde}, \citenamefont {Zhang},
  \citenamefont {Hu}, \citenamefont {Chen},\ and\ \citenamefont
  {Xiao}}]{BaiPRL2017}%
  \BibitemOpen
  \bibfield  {author} {\bibinfo {author} {\bibfnamefont {L.}~\bibnamefont
  {Bai}}, \bibinfo {author} {\bibfnamefont {M.}~\bibnamefont {Harder}},
  \bibinfo {author} {\bibfnamefont {P.}~\bibnamefont {Hyde}}, \bibinfo {author}
  {\bibfnamefont {Z.}~\bibnamefont {Zhang}}, \bibinfo {author} {\bibfnamefont
  {C.-M.}\ \bibnamefont {Hu}}, \bibinfo {author} {\bibfnamefont {Y.~P.}\
  \bibnamefont {Chen}}, \ and\ \bibinfo {author} {\bibfnamefont {J.~Q.}\
  \bibnamefont {Xiao}},\ }\href {\doibase 10.1103/PhysRevLett.118.217201}
  {\bibfield  {journal} {\bibinfo  {journal} {Phys. Rev. Lett.}\ }\textbf
  {\bibinfo {volume} {118}},\ \bibinfo {pages} {217201} (\bibinfo {year}
  {2017})}\BibitemShut {NoStop}%
\bibitem [{\citenamefont {Harder}\ \emph {et~al.}(2018)\citenamefont {Harder},
  \citenamefont {Yang}, \citenamefont {Yao}, \citenamefont {Yu}, \citenamefont
  {Rao}, \citenamefont {Gui}, \citenamefont {Stamps},\ and\ \citenamefont
  {Hu}}]{HarderPRL2018}%
  \BibitemOpen
  \bibfield  {author} {\bibinfo {author} {\bibfnamefont {M.}~\bibnamefont
  {Harder}}, \bibinfo {author} {\bibfnamefont {Y.}~\bibnamefont {Yang}},
  \bibinfo {author} {\bibfnamefont {B.~M.}\ \bibnamefont {Yao}}, \bibinfo
  {author} {\bibfnamefont {C.~H.}\ \bibnamefont {Yu}}, \bibinfo {author}
  {\bibfnamefont {J.~W.}\ \bibnamefont {Rao}}, \bibinfo {author} {\bibfnamefont
  {Y.~S.}\ \bibnamefont {Gui}}, \bibinfo {author} {\bibfnamefont {R.~L.}\
  \bibnamefont {Stamps}}, \ and\ \bibinfo {author} {\bibfnamefont {C.-M.}\
  \bibnamefont {Hu}},\ }\href {\doibase 10.1103/PhysRevLett.121.137203}
  {\bibfield  {journal} {\bibinfo  {journal} {Phys. Rev. Lett.}\ }\textbf
  {\bibinfo {volume} {121}},\ \bibinfo {pages} {137203} (\bibinfo {year}
  {2018})}\BibitemShut {NoStop}%
\bibitem [{\citenamefont {Li}\ \emph {et~al.}(2019{\natexlab{a}})\citenamefont
  {Li}, \citenamefont {Polakovic}, \citenamefont {Wang}, \citenamefont {Xu},
  \citenamefont {Lendinez}, \citenamefont {Zhang}, \citenamefont {Ding},
  \citenamefont {Khaire}, \citenamefont {Saglam}, \citenamefont {Divan},
  \citenamefont {Pearson}, \citenamefont {Kwok}, \citenamefont {Xiao},
  \citenamefont {Novosad}, \citenamefont {Hoffmann},\ and\ \citenamefont
  {Zhang}}]{LiPRL2019_magnon}%
  \BibitemOpen
  \bibfield  {author} {\bibinfo {author} {\bibfnamefont {Y.}~\bibnamefont
  {Li}}, \bibinfo {author} {\bibfnamefont {T.}~\bibnamefont {Polakovic}},
  \bibinfo {author} {\bibfnamefont {Y.-L.}\ \bibnamefont {Wang}}, \bibinfo
  {author} {\bibfnamefont {J.}~\bibnamefont {Xu}}, \bibinfo {author}
  {\bibfnamefont {S.}~\bibnamefont {Lendinez}}, \bibinfo {author}
  {\bibfnamefont {Z.}~\bibnamefont {Zhang}}, \bibinfo {author} {\bibfnamefont
  {J.}~\bibnamefont {Ding}}, \bibinfo {author} {\bibfnamefont {T.}~\bibnamefont
  {Khaire}}, \bibinfo {author} {\bibfnamefont {H.}~\bibnamefont {Saglam}},
  \bibinfo {author} {\bibfnamefont {R.}~\bibnamefont {Divan}}, \bibinfo
  {author} {\bibfnamefont {J.}~\bibnamefont {Pearson}}, \bibinfo {author}
  {\bibfnamefont {W.-K.}\ \bibnamefont {Kwok}}, \bibinfo {author}
  {\bibfnamefont {Z.}~\bibnamefont {Xiao}}, \bibinfo {author} {\bibfnamefont
  {V.}~\bibnamefont {Novosad}}, \bibinfo {author} {\bibfnamefont
  {A.}~\bibnamefont {Hoffmann}}, \ and\ \bibinfo {author} {\bibfnamefont
  {W.}~\bibnamefont {Zhang}},\ }\href {\doibase 10.1103/PhysRevLett.123.107701}
  {\bibfield  {journal} {\bibinfo  {journal} {Phys. Rev. Lett.}\ }\textbf
  {\bibinfo {volume} {123}},\ \bibinfo {pages} {107701} (\bibinfo {year}
  {2019}{\natexlab{a}})}\BibitemShut {NoStop}%
\bibitem [{\citenamefont {Hou}\ and\ \citenamefont {Liu}(2019)}]{HouPRL2019}%
  \BibitemOpen
  \bibfield  {author} {\bibinfo {author} {\bibfnamefont {J.~T.}\ \bibnamefont
  {Hou}}\ and\ \bibinfo {author} {\bibfnamefont {L.}~\bibnamefont {Liu}},\
  }\href {\doibase 10.1103/PhysRevLett.123.107702} {\bibfield  {journal}
  {\bibinfo  {journal} {Phys. Rev. Lett.}\ }\textbf {\bibinfo {volume} {123}},\
  \bibinfo {pages} {107702} (\bibinfo {year} {2019})}\BibitemShut {NoStop}%
\bibitem [{\citenamefont {McKenzie-Sell}\ \emph {et~al.}(2019)\citenamefont
  {McKenzie-Sell}, \citenamefont {Xie}, \citenamefont {Lee}, \citenamefont
  {Robinson}, \citenamefont {Ciccarelli},\ and\ \citenamefont
  {Haigh}}]{McKenziePRB2019}%
  \BibitemOpen
  \bibfield  {author} {\bibinfo {author} {\bibfnamefont {L.}~\bibnamefont
  {McKenzie-Sell}}, \bibinfo {author} {\bibfnamefont {J.}~\bibnamefont {Xie}},
  \bibinfo {author} {\bibfnamefont {C.-M.}\ \bibnamefont {Lee}}, \bibinfo
  {author} {\bibfnamefont {J.~W.~A.}\ \bibnamefont {Robinson}}, \bibinfo
  {author} {\bibfnamefont {C.}~\bibnamefont {Ciccarelli}}, \ and\ \bibinfo
  {author} {\bibfnamefont {J.~A.}\ \bibnamefont {Haigh}},\ }\href {\doibase
  10.1103/PhysRevB.99.140414} {\bibfield  {journal} {\bibinfo  {journal} {Phys.
  Rev. B}\ }\textbf {\bibinfo {volume} {99}},\ \bibinfo {pages} {140414}
  (\bibinfo {year} {2019})}\BibitemShut {NoStop}%
\bibitem [{\citenamefont {Lachance-Quirion}\ \emph {et~al.}(2020)\citenamefont
  {Lachance-Quirion}, \citenamefont {Wolski}, \citenamefont {Tabuchi},
  \citenamefont {Kono}, \citenamefont {Usami},\ and\ \citenamefont
  {Nakamura}}]{LachanceQuirionScience2020}%
  \BibitemOpen
  \bibfield  {author} {\bibinfo {author} {\bibfnamefont {D.}~\bibnamefont
  {Lachance-Quirion}}, \bibinfo {author} {\bibfnamefont {S.~P.}\ \bibnamefont
  {Wolski}}, \bibinfo {author} {\bibfnamefont {Y.}~\bibnamefont {Tabuchi}},
  \bibinfo {author} {\bibfnamefont {S.}~\bibnamefont {Kono}}, \bibinfo {author}
  {\bibfnamefont {K.}~\bibnamefont {Usami}}, \ and\ \bibinfo {author}
  {\bibfnamefont {Y.}~\bibnamefont {Nakamura}},\ }\href@noop {} {\bibfield
  {journal} {\bibinfo  {journal} {Science}\ }\textbf {\bibinfo {volume}
  {367}},\ \bibinfo {pages} {425} (\bibinfo {year} {2020})}\BibitemShut
  {NoStop}%
\bibitem [{\citenamefont {Osada}\ \emph {et~al.}(2016)\citenamefont {Osada},
  \citenamefont {Hisatomi}, \citenamefont {Noguchi}, \citenamefont {Tabuchi},
  \citenamefont {Yamazaki}, \citenamefont {Usami}, \citenamefont {Sadgrove},
  \citenamefont {Yalla}, \citenamefont {Nomura},\ and\ \citenamefont
  {Nakamura}}]{OsadaPRL2016}%
  \BibitemOpen
  \bibfield  {author} {\bibinfo {author} {\bibfnamefont {A.}~\bibnamefont
  {Osada}}, \bibinfo {author} {\bibfnamefont {R.}~\bibnamefont {Hisatomi}},
  \bibinfo {author} {\bibfnamefont {A.}~\bibnamefont {Noguchi}}, \bibinfo
  {author} {\bibfnamefont {Y.}~\bibnamefont {Tabuchi}}, \bibinfo {author}
  {\bibfnamefont {R.}~\bibnamefont {Yamazaki}}, \bibinfo {author}
  {\bibfnamefont {K.}~\bibnamefont {Usami}}, \bibinfo {author} {\bibfnamefont
  {M.}~\bibnamefont {Sadgrove}}, \bibinfo {author} {\bibfnamefont
  {R.}~\bibnamefont {Yalla}}, \bibinfo {author} {\bibfnamefont
  {M.}~\bibnamefont {Nomura}}, \ and\ \bibinfo {author} {\bibfnamefont
  {Y.}~\bibnamefont {Nakamura}},\ }\href {\doibase
  10.1103/PhysRevLett.116.223601} {\bibfield  {journal} {\bibinfo  {journal}
  {Phys. Rev. Lett.}\ }\textbf {\bibinfo {volume} {116}},\ \bibinfo {pages}
  {223601} (\bibinfo {year} {2016})}\BibitemShut {NoStop}%
\bibitem [{\citenamefont {Zhang}\ \emph
  {et~al.}(2016{\natexlab{a}})\citenamefont {Zhang}, \citenamefont {Zhu},
  \citenamefont {Zou},\ and\ \citenamefont {Tang}}]{ZhangPRL2016}%
  \BibitemOpen
  \bibfield  {author} {\bibinfo {author} {\bibfnamefont {X.}~\bibnamefont
  {Zhang}}, \bibinfo {author} {\bibfnamefont {N.}~\bibnamefont {Zhu}}, \bibinfo
  {author} {\bibfnamefont {C.-L.}\ \bibnamefont {Zou}}, \ and\ \bibinfo
  {author} {\bibfnamefont {H.~X.}\ \bibnamefont {Tang}},\ }\href {\doibase
  10.1103/PhysRevLett.117.123605} {\bibfield  {journal} {\bibinfo  {journal}
  {Phys. Rev. Lett.}\ }\textbf {\bibinfo {volume} {117}},\ \bibinfo {pages}
  {123605} (\bibinfo {year} {2016}{\natexlab{a}})}\BibitemShut {NoStop}%
\bibitem [{\citenamefont {Zhang}\ \emph
  {et~al.}(2016{\natexlab{b}})\citenamefont {Zhang}, \citenamefont {Zou},
  \citenamefont {Jiang},\ and\ \citenamefont {Tang}}]{ZhangScienceAdv2016}%
  \BibitemOpen
  \bibfield  {author} {\bibinfo {author} {\bibfnamefont {X.}~\bibnamefont
  {Zhang}}, \bibinfo {author} {\bibfnamefont {C.-L.}\ \bibnamefont {Zou}},
  \bibinfo {author} {\bibfnamefont {L.}~\bibnamefont {Jiang}}, \ and\ \bibinfo
  {author} {\bibfnamefont {H.~X.}\ \bibnamefont {Tang}},\ }\href@noop {}
  {\bibfield  {journal} {\bibinfo  {journal} {Sci. Adv.}\ }\textbf {\bibinfo
  {volume} {2}},\ \bibinfo {pages} {e1501286} (\bibinfo {year}
  {2016}{\natexlab{b}})}\BibitemShut {NoStop}%
\bibitem [{\citenamefont {An}\ \emph {et~al.}(2020)\citenamefont {An},
  \citenamefont {Litvinenko}, \citenamefont {Kohno}, \citenamefont {Fuad},
  \citenamefont {Naletov}, \citenamefont {Vila}, \citenamefont {Ebels},
  \citenamefont {de~Loubens}, \citenamefont {Hurdequint}, \citenamefont
  {Beaulieu}, \citenamefont {Ben~Youssef}, \citenamefont {Vukadinovic},
  \citenamefont {Bauer}, \citenamefont {Slavin}, \citenamefont {Tiberkevich},\
  and\ \citenamefont {Klein}}]{AnPRB2020}%
  \BibitemOpen
  \bibfield  {author} {\bibinfo {author} {\bibfnamefont {K.}~\bibnamefont
  {An}}, \bibinfo {author} {\bibfnamefont {A.~N.}\ \bibnamefont {Litvinenko}},
  \bibinfo {author} {\bibfnamefont {R.}~\bibnamefont {Kohno}}, \bibinfo
  {author} {\bibfnamefont {A.~A.}\ \bibnamefont {Fuad}}, \bibinfo {author}
  {\bibfnamefont {V.~V.}\ \bibnamefont {Naletov}}, \bibinfo {author}
  {\bibfnamefont {L.}~\bibnamefont {Vila}}, \bibinfo {author} {\bibfnamefont
  {U.}~\bibnamefont {Ebels}}, \bibinfo {author} {\bibfnamefont
  {G.}~\bibnamefont {de~Loubens}}, \bibinfo {author} {\bibfnamefont
  {H.}~\bibnamefont {Hurdequint}}, \bibinfo {author} {\bibfnamefont
  {N.}~\bibnamefont {Beaulieu}}, \bibinfo {author} {\bibfnamefont
  {J.}~\bibnamefont {Ben~Youssef}}, \bibinfo {author} {\bibfnamefont
  {N.}~\bibnamefont {Vukadinovic}}, \bibinfo {author} {\bibfnamefont
  {G.~E.~W.}\ \bibnamefont {Bauer}}, \bibinfo {author} {\bibfnamefont {A.~N.}\
  \bibnamefont {Slavin}}, \bibinfo {author} {\bibfnamefont {V.~S.}\
  \bibnamefont {Tiberkevich}}, \ and\ \bibinfo {author} {\bibfnamefont
  {O.}~\bibnamefont {Klein}},\ }\href {\doibase 10.1103/PhysRevB.101.060407}
  {\bibfield  {journal} {\bibinfo  {journal} {Phys. Rev. B}\ }\textbf {\bibinfo
  {volume} {101}},\ \bibinfo {pages} {060407} (\bibinfo {year}
  {2020})}\BibitemShut {NoStop}%
\bibitem [{\citenamefont {Zhao}\ \emph {et~al.}(2020)\citenamefont {Zhao},
  \citenamefont {Li}, \citenamefont {Zhang}, \citenamefont {Vogel},
  \citenamefont {Pearson}, \citenamefont {Wang}, \citenamefont {Zhang},
  \citenamefont {Novosad}, \citenamefont {Liu},\ and\ \citenamefont
  {Hoffmann}}]{ZhaoChenboPRApplied2020}%
  \BibitemOpen
  \bibfield  {author} {\bibinfo {author} {\bibfnamefont {C.}~\bibnamefont
  {Zhao}}, \bibinfo {author} {\bibfnamefont {Y.}~\bibnamefont {Li}}, \bibinfo
  {author} {\bibfnamefont {Z.}~\bibnamefont {Zhang}}, \bibinfo {author}
  {\bibfnamefont {M.}~\bibnamefont {Vogel}}, \bibinfo {author} {\bibfnamefont
  {J.~E.}\ \bibnamefont {Pearson}}, \bibinfo {author} {\bibfnamefont
  {J.}~\bibnamefont {Wang}}, \bibinfo {author} {\bibfnamefont {W.}~\bibnamefont
  {Zhang}}, \bibinfo {author} {\bibfnamefont {V.}~\bibnamefont {Novosad}},
  \bibinfo {author} {\bibfnamefont {Q.}~\bibnamefont {Liu}}, \ and\ \bibinfo
  {author} {\bibfnamefont {A.}~\bibnamefont {Hoffmann}},\ }\href@noop {}
  {\bibfield  {journal} {\bibinfo  {journal} {Phys. Rev. Applied}\ }\textbf
  {\bibinfo {volume} {13}},\ \bibinfo {pages} {054032} (\bibinfo {year}
  {2020})}\BibitemShut {NoStop}%
\bibitem [{\citenamefont {Klingler}\ \emph {et~al.}(2018)\citenamefont
  {Klingler}, \citenamefont {Amin}, \citenamefont {Gepr\"ags}, \citenamefont
  {Ganzhorn}, \citenamefont {Maier-Flaig}, \citenamefont {Althammer},
  \citenamefont {Huebl}, \citenamefont {Gross}, \citenamefont {McMichael},
  \citenamefont {Stiles}, \citenamefont {Goennenwein},\ and\ \citenamefont
  {Weiler}}]{KlinglerPRL2018}%
  \BibitemOpen
  \bibfield  {author} {\bibinfo {author} {\bibfnamefont {S.}~\bibnamefont
  {Klingler}}, \bibinfo {author} {\bibfnamefont {V.}~\bibnamefont {Amin}},
  \bibinfo {author} {\bibfnamefont {S.}~\bibnamefont {Gepr\"ags}}, \bibinfo
  {author} {\bibfnamefont {K.}~\bibnamefont {Ganzhorn}}, \bibinfo {author}
  {\bibfnamefont {H.}~\bibnamefont {Maier-Flaig}}, \bibinfo {author}
  {\bibfnamefont {M.}~\bibnamefont {Althammer}}, \bibinfo {author}
  {\bibfnamefont {H.}~\bibnamefont {Huebl}}, \bibinfo {author} {\bibfnamefont
  {R.}~\bibnamefont {Gross}}, \bibinfo {author} {\bibfnamefont {R.~D.}\
  \bibnamefont {McMichael}}, \bibinfo {author} {\bibfnamefont {M.~D.}\
  \bibnamefont {Stiles}}, \bibinfo {author} {\bibfnamefont {S.~T.~B.}\
  \bibnamefont {Goennenwein}}, \ and\ \bibinfo {author} {\bibfnamefont
  {M.}~\bibnamefont {Weiler}},\ }\href {\doibase
  10.1103/PhysRevLett.120.127201} {\bibfield  {journal} {\bibinfo  {journal}
  {Phys. Rev. Lett.}\ }\textbf {\bibinfo {volume} {120}},\ \bibinfo {pages}
  {127201} (\bibinfo {year} {2018})}\BibitemShut {NoStop}%
\bibitem [{\citenamefont {Chen}\ \emph {et~al.}(2018)\citenamefont {Chen},
  \citenamefont {Liu}, \citenamefont {Liu}, \citenamefont {Xiao}, \citenamefont
  {Xia}, \citenamefont {Bauer}, \citenamefont {Wu},\ and\ \citenamefont
  {Yu}}]{ChenPRL2018}%
  \BibitemOpen
  \bibfield  {author} {\bibinfo {author} {\bibfnamefont {J.}~\bibnamefont
  {Chen}}, \bibinfo {author} {\bibfnamefont {C.}~\bibnamefont {Liu}}, \bibinfo
  {author} {\bibfnamefont {T.}~\bibnamefont {Liu}}, \bibinfo {author}
  {\bibfnamefont {Y.}~\bibnamefont {Xiao}}, \bibinfo {author} {\bibfnamefont
  {K.}~\bibnamefont {Xia}}, \bibinfo {author} {\bibfnamefont {G.~E.~W.}\
  \bibnamefont {Bauer}}, \bibinfo {author} {\bibfnamefont {M.}~\bibnamefont
  {Wu}}, \ and\ \bibinfo {author} {\bibfnamefont {H.}~\bibnamefont {Yu}},\
  }\href {\doibase 10.1103/PhysRevLett.120.217202} {\bibfield  {journal}
  {\bibinfo  {journal} {Phys. Rev. Lett.}\ }\textbf {\bibinfo {volume} {120}},\
  \bibinfo {pages} {217202} (\bibinfo {year} {2018})}\BibitemShut {NoStop}%
\bibitem [{\citenamefont {Qin}\ \emph {et~al.}(2018)\citenamefont {Qin},
  \citenamefont {H\"{a}m\"{a}l\"{a}inen},\ and\ \citenamefont {van
  Dijken}}]{QinSREP2018}%
  \BibitemOpen
  \bibfield  {author} {\bibinfo {author} {\bibfnamefont {H.}~\bibnamefont
  {Qin}}, \bibinfo {author} {\bibfnamefont {S.~J.}\ \bibnamefont
  {H\"{a}m\"{a}l\"{a}inen}}, \ and\ \bibinfo {author} {\bibfnamefont
  {S.}~\bibnamefont {van Dijken}},\ }\href@noop {} {\bibfield  {journal}
  {\bibinfo  {journal} {Sci. Rep.}\ }\textbf {\bibinfo {volume} {8}},\ \bibinfo
  {pages} {5755} (\bibinfo {year} {2018})}\BibitemShut {NoStop}%
\bibitem [{\citenamefont {Li}\ \emph {et~al.}(2020{\natexlab{a}})\citenamefont
  {Li}, \citenamefont {Cao}, \citenamefont {Amin}, \citenamefont {Zhang},
  \citenamefont {Gibbons}, \citenamefont {Sklenar}, \citenamefont {Pearson},
  \citenamefont {Haney}, \citenamefont {Stiles}, \citenamefont {Bailey},
  \citenamefont {Novosad}, \citenamefont {Hoffmann},\ and\ \citenamefont
  {Zhang}}]{LiPRL2020_YIGPy}%
  \BibitemOpen
  \bibfield  {author} {\bibinfo {author} {\bibfnamefont {Y.}~\bibnamefont
  {Li}}, \bibinfo {author} {\bibfnamefont {W.}~\bibnamefont {Cao}}, \bibinfo
  {author} {\bibfnamefont {V.~P.}\ \bibnamefont {Amin}}, \bibinfo {author}
  {\bibfnamefont {Z.}~\bibnamefont {Zhang}}, \bibinfo {author} {\bibfnamefont
  {J.}~\bibnamefont {Gibbons}}, \bibinfo {author} {\bibfnamefont
  {J.}~\bibnamefont {Sklenar}}, \bibinfo {author} {\bibfnamefont
  {J.}~\bibnamefont {Pearson}}, \bibinfo {author} {\bibfnamefont {P.~M.}\
  \bibnamefont {Haney}}, \bibinfo {author} {\bibfnamefont {M.~D.}\ \bibnamefont
  {Stiles}}, \bibinfo {author} {\bibfnamefont {W.~E.}\ \bibnamefont {Bailey}},
  \bibinfo {author} {\bibfnamefont {V.}~\bibnamefont {Novosad}}, \bibinfo
  {author} {\bibfnamefont {A.}~\bibnamefont {Hoffmann}}, \ and\ \bibinfo
  {author} {\bibfnamefont {W.}~\bibnamefont {Zhang}},\ }\href@noop {}
  {\bibfield  {journal} {\bibinfo  {journal} {Phys. Rev. Lett.}\ }\textbf
  {\bibinfo {volume} {124}},\ \bibinfo {pages} {117202} (\bibinfo {year}
  {2020}{\natexlab{a}})}\BibitemShut {NoStop}%
\bibitem [{\citenamefont {Fan}\ \emph {et~al.}(2020)\citenamefont {Fan},
  \citenamefont {Quarterman}, \citenamefont {Finley}, \citenamefont {Han},
  \citenamefont {Zhang}, \citenamefont {Hou}, \citenamefont {Stiles},
  \citenamefont {Grutter},\ and\ \citenamefont {Liu}}]{FanPRApplied2020}%
  \BibitemOpen
  \bibfield  {author} {\bibinfo {author} {\bibfnamefont {Y.}~\bibnamefont
  {Fan}}, \bibinfo {author} {\bibfnamefont {P.}~\bibnamefont {Quarterman}},
  \bibinfo {author} {\bibfnamefont {J.}~\bibnamefont {Finley}}, \bibinfo
  {author} {\bibfnamefont {J.}~\bibnamefont {Han}}, \bibinfo {author}
  {\bibfnamefont {P.}~\bibnamefont {Zhang}}, \bibinfo {author} {\bibfnamefont
  {J.~T.}\ \bibnamefont {Hou}}, \bibinfo {author} {\bibfnamefont {M.~D.}\
  \bibnamefont {Stiles}}, \bibinfo {author} {\bibfnamefont {A.~J.}\
  \bibnamefont {Grutter}}, \ and\ \bibinfo {author} {\bibfnamefont
  {L.}~\bibnamefont {Liu}},\ }\href@noop {} {\bibfield  {journal} {\bibinfo
  {journal} {Phys. Rev. Applied}\ }\textbf {\bibinfo {volume} {13}},\ \bibinfo
  {pages} {061002} (\bibinfo {year} {2020})}\BibitemShut {NoStop}%
\bibitem [{\citenamefont {Xiong}\ \emph
  {et~al.}(2020{\natexlab{a}})\citenamefont {Xiong}, \citenamefont {Li},
  \citenamefont {Hammami}, \citenamefont {Bidthanapally}, \citenamefont
  {Sklenar}, \citenamefont {Zhang}, \citenamefont {Qu}, \citenamefont
  {Srinivasan}, \citenamefont {Pearson}, \citenamefont {Hoffmann},
  \citenamefont {Novosad},\ and\ \citenamefont {Zhang}}]{XiongSciRep2020}%
  \BibitemOpen
  \bibfield  {author} {\bibinfo {author} {\bibfnamefont {Y.}~\bibnamefont
  {Xiong}}, \bibinfo {author} {\bibfnamefont {Y.}~\bibnamefont {Li}}, \bibinfo
  {author} {\bibfnamefont {M.}~\bibnamefont {Hammami}}, \bibinfo {author}
  {\bibfnamefont {R.}~\bibnamefont {Bidthanapally}}, \bibinfo {author}
  {\bibfnamefont {J.}~\bibnamefont {Sklenar}}, \bibinfo {author} {\bibfnamefont
  {X.}~\bibnamefont {Zhang}}, \bibinfo {author} {\bibfnamefont
  {H.}~\bibnamefont {Qu}}, \bibinfo {author} {\bibfnamefont {G.}~\bibnamefont
  {Srinivasan}}, \bibinfo {author} {\bibfnamefont {J.}~\bibnamefont {Pearson}},
  \bibinfo {author} {\bibfnamefont {A.}~\bibnamefont {Hoffmann}}, \bibinfo
  {author} {\bibfnamefont {V.}~\bibnamefont {Novosad}}, \ and\ \bibinfo
  {author} {\bibfnamefont {W.}~\bibnamefont {Zhang}},\ }\href@noop {}
  {\bibfield  {journal} {\bibinfo  {journal} {Sci. Rep.}\ }\textbf {\bibinfo
  {volume} {10}},\ \bibinfo {pages} {12548} (\bibinfo {year}
  {2020}{\natexlab{a}})}\BibitemShut {NoStop}%
\bibitem [{\citenamefont {Xiong}\ \emph
  {et~al.}(2020{\natexlab{b}})\citenamefont {Xiong}, \citenamefont {Li},
  \citenamefont {Bidthanapally}, \citenamefont {Sklenar}, \citenamefont
  {Hammami}, \citenamefont {Hall}, \citenamefont {Zhang}, \citenamefont {Li},
  \citenamefont {Pearson}, \citenamefont {Sebastian}, \citenamefont
  {Srinivasan}, \citenamefont {Hoffmann}, \citenamefont {Qu}, \citenamefont
  {Novosad},\ and\ \citenamefont {Zhang}}]{XiongIEEE2020}%
  \BibitemOpen
  \bibfield  {author} {\bibinfo {author} {\bibfnamefont {Y.}~\bibnamefont
  {Xiong}}, \bibinfo {author} {\bibfnamefont {Y.}~\bibnamefont {Li}}, \bibinfo
  {author} {\bibfnamefont {R.}~\bibnamefont {Bidthanapally}}, \bibinfo {author}
  {\bibfnamefont {J.}~\bibnamefont {Sklenar}}, \bibinfo {author} {\bibfnamefont
  {M.}~\bibnamefont {Hammami}}, \bibinfo {author} {\bibfnamefont
  {S.}~\bibnamefont {Hall}}, \bibinfo {author} {\bibfnamefont {X.}~\bibnamefont
  {Zhang}}, \bibinfo {author} {\bibfnamefont {P.}~\bibnamefont {Li}}, \bibinfo
  {author} {\bibfnamefont {J.~E.}\ \bibnamefont {Pearson}}, \bibinfo {author}
  {\bibfnamefont {T.}~\bibnamefont {Sebastian}}, \bibinfo {author}
  {\bibfnamefont {G.}~\bibnamefont {Srinivasan}}, \bibinfo {author}
  {\bibfnamefont {A.}~\bibnamefont {Hoffmann}}, \bibinfo {author}
  {\bibfnamefont {H.}~\bibnamefont {Qu}}, \bibinfo {author} {\bibfnamefont
  {V.}~\bibnamefont {Novosad}}, \ and\ \bibinfo {author} {\bibfnamefont
  {W.}~\bibnamefont {Zhang}},\ }\href {\doibase 10.1109/TMAG.2020.3013063}
  {\bibfield  {journal} {\bibinfo  {journal} {IEEE Trans. Magn.}\ } (\bibinfo
  {year} {2020}{\natexlab{b}}),\ 10.1109/TMAG.2020.3013063}\BibitemShut
  {NoStop}%
\bibitem [{\citenamefont {MacNeill}\ \emph {et~al.}(2019)\citenamefont
  {MacNeill}, \citenamefont {Hou}, \citenamefont {Klein}, \citenamefont
  {Zhang}, \citenamefont {Jarillo-Herrero},\ and\ \citenamefont
  {Liu}}]{MacNeillPRL2019}%
  \BibitemOpen
  \bibfield  {author} {\bibinfo {author} {\bibfnamefont {D.}~\bibnamefont
  {MacNeill}}, \bibinfo {author} {\bibfnamefont {J.~T.}\ \bibnamefont {Hou}},
  \bibinfo {author} {\bibfnamefont {D.~R.}\ \bibnamefont {Klein}}, \bibinfo
  {author} {\bibfnamefont {P.}~\bibnamefont {Zhang}}, \bibinfo {author}
  {\bibfnamefont {P.}~\bibnamefont {Jarillo-Herrero}}, \ and\ \bibinfo {author}
  {\bibfnamefont {L.}~\bibnamefont {Liu}},\ }\href {\doibase
  10.1103/PhysRevLett.123.047204} {\bibfield  {journal} {\bibinfo  {journal}
  {Phys. Rev. Lett.}\ }\textbf {\bibinfo {volume} {123}},\ \bibinfo {pages}
  {047204} (\bibinfo {year} {2019})}\BibitemShut {NoStop}%
\bibitem [{\citenamefont {Liensberger}\ \emph {et~al.}(2019)\citenamefont
  {Liensberger}, \citenamefont {Kamra}, \citenamefont {Maier-Flaig},
  \citenamefont {Gepr\"ags}, \citenamefont {Erb}, \citenamefont {Goennenwein},
  \citenamefont {Gross}, \citenamefont {Belzig}, \citenamefont {Huebl},\ and\
  \citenamefont {Weiler}}]{LiensbergerPRL2019}%
  \BibitemOpen
  \bibfield  {author} {\bibinfo {author} {\bibfnamefont {L.}~\bibnamefont
  {Liensberger}}, \bibinfo {author} {\bibfnamefont {A.}~\bibnamefont {Kamra}},
  \bibinfo {author} {\bibfnamefont {H.}~\bibnamefont {Maier-Flaig}}, \bibinfo
  {author} {\bibfnamefont {S.}~\bibnamefont {Gepr\"ags}}, \bibinfo {author}
  {\bibfnamefont {A.}~\bibnamefont {Erb}}, \bibinfo {author} {\bibfnamefont
  {S.~T.~B.}\ \bibnamefont {Goennenwein}}, \bibinfo {author} {\bibfnamefont
  {R.}~\bibnamefont {Gross}}, \bibinfo {author} {\bibfnamefont
  {W.}~\bibnamefont {Belzig}}, \bibinfo {author} {\bibfnamefont
  {H.}~\bibnamefont {Huebl}}, \ and\ \bibinfo {author} {\bibfnamefont
  {M.}~\bibnamefont {Weiler}},\ }\href {\doibase
  10.1103/PhysRevLett.123.117204} {\bibfield  {journal} {\bibinfo  {journal}
  {Phys. Rev. Lett.}\ }\textbf {\bibinfo {volume} {123}},\ \bibinfo {pages}
  {117204} (\bibinfo {year} {2019})}\BibitemShut {NoStop}%
\bibitem [{\citenamefont {Dai}\ \emph {et~al.}(2020)\citenamefont {Dai},
  \citenamefont {Xie}, \citenamefont {Pan},\ and\ \citenamefont
  {Ma}}]{DaiJAP2020}%
  \BibitemOpen
  \bibfield  {author} {\bibinfo {author} {\bibfnamefont {C.}~\bibnamefont
  {Dai}}, \bibinfo {author} {\bibfnamefont {K.}~\bibnamefont {Xie}}, \bibinfo
  {author} {\bibfnamefont {Z.}~\bibnamefont {Pan}}, \ and\ \bibinfo {author}
  {\bibfnamefont {F.}~\bibnamefont {Ma}},\ }\href@noop {} {\bibfield  {journal}
  {\bibinfo  {journal} {J. Appl. Phys.}\ }\textbf {\bibinfo {volume} {127}},\
  \bibinfo {pages} {203902} (\bibinfo {year} {2020})}\BibitemShut {NoStop}%
\bibitem [{\citenamefont {Zhang}\ \emph {et~al.}(2017)\citenamefont {Zhang},
  \citenamefont {Luo}, \citenamefont {Wang}, \citenamefont {Li},\ and\
  \citenamefont {You}}]{ZhangDengkeNComm2017}%
  \BibitemOpen
  \bibfield  {author} {\bibinfo {author} {\bibfnamefont {D.}~\bibnamefont
  {Zhang}}, \bibinfo {author} {\bibfnamefont {X.-Q.}\ \bibnamefont {Luo}},
  \bibinfo {author} {\bibfnamefont {Y.-P.}\ \bibnamefont {Wang}}, \bibinfo
  {author} {\bibfnamefont {T.-F.}\ \bibnamefont {Li}}, \ and\ \bibinfo {author}
  {\bibfnamefont {J.}~\bibnamefont {You}},\ }\href@noop {} {\bibfield
  {journal} {\bibinfo  {journal} {Nature Commun.}\ }\textbf {\bibinfo {volume}
  {8}},\ \bibinfo {pages} {1368} (\bibinfo {year} {2017})}\BibitemShut
  {NoStop}%
\bibitem [{\citenamefont {Zhang}\ \emph {et~al.}(2019)\citenamefont {Zhang},
  \citenamefont {Ding}, \citenamefont {Zhou}, \citenamefont {Xu},\ and\
  \citenamefont {Jin}}]{ZhangPRL2019}%
  \BibitemOpen
  \bibfield  {author} {\bibinfo {author} {\bibfnamefont {X.}~\bibnamefont
  {Zhang}}, \bibinfo {author} {\bibfnamefont {K.}~\bibnamefont {Ding}},
  \bibinfo {author} {\bibfnamefont {X.}~\bibnamefont {Zhou}}, \bibinfo {author}
  {\bibfnamefont {J.}~\bibnamefont {Xu}}, \ and\ \bibinfo {author}
  {\bibfnamefont {D.}~\bibnamefont {Jin}},\ }\href {\doibase
  10.1103/PhysRevLett.123.237202} {\bibfield  {journal} {\bibinfo  {journal}
  {Phys. Rev. Lett.}\ }\textbf {\bibinfo {volume} {123}},\ \bibinfo {pages}
  {237202} (\bibinfo {year} {2019})}\BibitemShut {NoStop}%
\bibitem [{\citenamefont {Bhoi}\ \emph {et~al.}(2019)\citenamefont {Bhoi},
  \citenamefont {Kim}, \citenamefont {Jang}, \citenamefont {Kim}, \citenamefont
  {Yang}, \citenamefont {Cho},\ and\ \citenamefont {Kim}}]{BhoiPRB2019}%
  \BibitemOpen
  \bibfield  {author} {\bibinfo {author} {\bibfnamefont {B.}~\bibnamefont
  {Bhoi}}, \bibinfo {author} {\bibfnamefont {B.}~\bibnamefont {Kim}}, \bibinfo
  {author} {\bibfnamefont {S.-H.}\ \bibnamefont {Jang}}, \bibinfo {author}
  {\bibfnamefont {J.}~\bibnamefont {Kim}}, \bibinfo {author} {\bibfnamefont
  {J.}~\bibnamefont {Yang}}, \bibinfo {author} {\bibfnamefont {Y.-J.}\
  \bibnamefont {Cho}}, \ and\ \bibinfo {author} {\bibfnamefont {S.-K.}\
  \bibnamefont {Kim}},\ }\href {\doibase 10.1103/PhysRevB.99.134426} {\bibfield
   {journal} {\bibinfo  {journal} {Phys. Rev. B}\ }\textbf {\bibinfo {volume}
  {99}},\ \bibinfo {pages} {134426} (\bibinfo {year} {2019})}\BibitemShut
  {NoStop}%
\bibitem [{\citenamefont {Boventer}\ \emph {et~al.}(2020)\citenamefont
  {Boventer}, \citenamefont {D\"orflinger}, \citenamefont {Wolz}, \citenamefont
  {Mac\^edo}, \citenamefont {Lebrun}, \citenamefont {Kl\"aui},\ and\
  \citenamefont {Weides}}]{BoventerPRResearch2020}%
  \BibitemOpen
  \bibfield  {author} {\bibinfo {author} {\bibfnamefont {I.}~\bibnamefont
  {Boventer}}, \bibinfo {author} {\bibfnamefont {C.}~\bibnamefont
  {D\"orflinger}}, \bibinfo {author} {\bibfnamefont {T.}~\bibnamefont {Wolz}},
  \bibinfo {author} {\bibfnamefont {R.}~\bibnamefont {Mac\^edo}}, \bibinfo
  {author} {\bibfnamefont {R.}~\bibnamefont {Lebrun}}, \bibinfo {author}
  {\bibfnamefont {M.}~\bibnamefont {Kl\"aui}}, \ and\ \bibinfo {author}
  {\bibfnamefont {M.}~\bibnamefont {Weides}},\ }\href {\doibase
  10.1103/PhysRevResearch.2.013154} {\bibfield  {journal} {\bibinfo  {journal}
  {Phys. Rev. Research}\ }\textbf {\bibinfo {volume} {2}},\ \bibinfo {pages}
  {013154} (\bibinfo {year} {2020})}\BibitemShut {NoStop}%
\bibitem [{\citenamefont {Wang}\ \emph
  {et~al.}(2019{\natexlab{a}})\citenamefont {Wang}, \citenamefont {Rao},
  \citenamefont {Yang}, \citenamefont {Xu}, \citenamefont {Gui}, \citenamefont
  {Yao}, \citenamefont {You},\ and\ \citenamefont {Hu}}]{WangPRL2019}%
  \BibitemOpen
  \bibfield  {author} {\bibinfo {author} {\bibfnamefont {Y.-P.}\ \bibnamefont
  {Wang}}, \bibinfo {author} {\bibfnamefont {J.~W.}\ \bibnamefont {Rao}},
  \bibinfo {author} {\bibfnamefont {Y.}~\bibnamefont {Yang}}, \bibinfo {author}
  {\bibfnamefont {P.-C.}\ \bibnamefont {Xu}}, \bibinfo {author} {\bibfnamefont
  {Y.~S.}\ \bibnamefont {Gui}}, \bibinfo {author} {\bibfnamefont {B.~M.}\
  \bibnamefont {Yao}}, \bibinfo {author} {\bibfnamefont {J.~Q.}\ \bibnamefont
  {You}}, \ and\ \bibinfo {author} {\bibfnamefont {C.-M.}\ \bibnamefont {Hu}},\
  }\href {\doibase 10.1103/PhysRevLett.123.127202} {\bibfield  {journal}
  {\bibinfo  {journal} {Phys. Rev. Lett.}\ }\textbf {\bibinfo {volume} {123}},\
  \bibinfo {pages} {127202} (\bibinfo {year} {2019}{\natexlab{a}})}\BibitemShut
  {NoStop}%
\bibitem [{\citenamefont {Zhang}\ \emph {et~al.}(2020)\citenamefont {Zhang},
  \citenamefont {Galda}, \citenamefont {Han}, \citenamefont {Jin},\ and\
  \citenamefont {Vinokur}}]{ZhangPRApplied2020}%
  \BibitemOpen
  \bibfield  {author} {\bibinfo {author} {\bibfnamefont {X.}~\bibnamefont
  {Zhang}}, \bibinfo {author} {\bibfnamefont {A.}~\bibnamefont {Galda}},
  \bibinfo {author} {\bibfnamefont {X.}~\bibnamefont {Han}}, \bibinfo {author}
  {\bibfnamefont {D.}~\bibnamefont {Jin}}, \ and\ \bibinfo {author}
  {\bibfnamefont {V.~M.}\ \bibnamefont {Vinokur}},\ }\href {\doibase
  10.1103/PhysRevApplied.13.044039} {\bibfield  {journal} {\bibinfo  {journal}
  {Phys. Rev. Applied}\ }\textbf {\bibinfo {volume} {13}},\ \bibinfo {pages}
  {044039} (\bibinfo {year} {2020})}\BibitemShut {NoStop}%
\bibitem [{\citenamefont {Juretschke}(1960)}]{JuretschkeJAP1960}%
  \BibitemOpen
  \bibfield  {author} {\bibinfo {author} {\bibfnamefont {H.~J.}\ \bibnamefont
  {Juretschke}},\ }\href@noop {} {\bibfield  {journal} {\bibinfo  {journal} {J.
  Appl. Phys.}\ }\textbf {\bibinfo {volume} {31}},\ \bibinfo {pages} {1401}
  (\bibinfo {year} {1960})}\BibitemShut {NoStop}%
\bibitem [{\citenamefont {Tulapurkar}\ \emph {et~al.}(2005)\citenamefont
  {Tulapurkar}, \citenamefont {Suzuki}, \citenamefont {Fukushima},
  \citenamefont {Kubota}, \citenamefont {Maehara}, \citenamefont {Tsunekawa},
  \citenamefont {Djayaprawira}, \citenamefont {Watanabe},\ and\ \citenamefont
  {Yuasa}}]{TulapurkarNature2005}%
  \BibitemOpen
  \bibfield  {author} {\bibinfo {author} {\bibfnamefont {A.~A.}\ \bibnamefont
  {Tulapurkar}}, \bibinfo {author} {\bibfnamefont {Y.}~\bibnamefont {Suzuki}},
  \bibinfo {author} {\bibfnamefont {A.}~\bibnamefont {Fukushima}}, \bibinfo
  {author} {\bibfnamefont {H.}~\bibnamefont {Kubota}}, \bibinfo {author}
  {\bibfnamefont {H.}~\bibnamefont {Maehara}}, \bibinfo {author} {\bibfnamefont
  {K.}~\bibnamefont {Tsunekawa}}, \bibinfo {author} {\bibfnamefont {D.~D.}\
  \bibnamefont {Djayaprawira}}, \bibinfo {author} {\bibfnamefont
  {N.}~\bibnamefont {Watanabe}}, \ and\ \bibinfo {author} {\bibfnamefont
  {S.}~\bibnamefont {Yuasa}},\ }\href@noop {} {\bibfield  {journal} {\bibinfo
  {journal} {Nature}\ }\textbf {\bibinfo {volume} {438}},\ \bibinfo {pages}
  {339} (\bibinfo {year} {2005})}\BibitemShut {NoStop}%
\bibitem [{\citenamefont {Fuchs}\ \emph {et~al.}(2007)\citenamefont {Fuchs},
  \citenamefont {Sankey}, \citenamefont {Pribiag}, \citenamefont {Qian},
  \citenamefont {Braganca}, \citenamefont {Garcia}, \citenamefont {Ryan},
  \citenamefont {Li}, \citenamefont {Ozatay}, \citenamefont {Ralph},\ and\
  \citenamefont {Buhrman}}]{FuchsAPL2007}%
  \BibitemOpen
  \bibfield  {author} {\bibinfo {author} {\bibfnamefont {G.~D.}\ \bibnamefont
  {Fuchs}}, \bibinfo {author} {\bibfnamefont {J.~C.}\ \bibnamefont {Sankey}},
  \bibinfo {author} {\bibfnamefont {V.~S.}\ \bibnamefont {Pribiag}}, \bibinfo
  {author} {\bibfnamefont {L.}~\bibnamefont {Qian}}, \bibinfo {author}
  {\bibfnamefont {P.~M.}\ \bibnamefont {Braganca}}, \bibinfo {author}
  {\bibfnamefont {A.~G.~F.}\ \bibnamefont {Garcia}}, \bibinfo {author}
  {\bibfnamefont {E.~M.}\ \bibnamefont {Ryan}}, \bibinfo {author}
  {\bibfnamefont {Z.-P.}\ \bibnamefont {Li}}, \bibinfo {author} {\bibfnamefont
  {O.}~\bibnamefont {Ozatay}}, \bibinfo {author} {\bibfnamefont {D.~C.}\
  \bibnamefont {Ralph}}, \ and\ \bibinfo {author} {\bibfnamefont {R.~A.}\
  \bibnamefont {Buhrman}},\ }\href@noop {} {\bibfield  {journal} {\bibinfo
  {journal} {Appl. Phys. Lett.}\ }\textbf {\bibinfo {volume} {91}},\ \bibinfo
  {pages} {062507} (\bibinfo {year} {2007})}\BibitemShut {NoStop}%
\bibitem [{\citenamefont {Gui}\ \emph {et~al.}(2007)\citenamefont {Gui},
  \citenamefont {Mecking}, \citenamefont {Zhou}, \citenamefont {Williams},\
  and\ \citenamefont {Hu}}]{GuiPRL2007}%
  \BibitemOpen
  \bibfield  {author} {\bibinfo {author} {\bibfnamefont {Y.~S.}\ \bibnamefont
  {Gui}}, \bibinfo {author} {\bibfnamefont {N.}~\bibnamefont {Mecking}},
  \bibinfo {author} {\bibfnamefont {X.}~\bibnamefont {Zhou}}, \bibinfo {author}
  {\bibfnamefont {G.}~\bibnamefont {Williams}}, \ and\ \bibinfo {author}
  {\bibfnamefont {C.-M.}\ \bibnamefont {Hu}},\ }\href@noop {} {\bibfield
  {journal} {\bibinfo  {journal} {Phys. Rev. Lett.}\ }\textbf {\bibinfo
  {volume} {98}},\ \bibinfo {pages} {107602} (\bibinfo {year}
  {2007})}\BibitemShut {NoStop}%
\bibitem [{\citenamefont {Sankey}\ \emph {et~al.}(2008)\citenamefont {Sankey},
  \citenamefont {Cui}, \citenamefont {Sun}, \citenamefont {Slonczewski},
  \citenamefont {Buhrman},\ and\ \citenamefont {Ralph}}]{SankeyNphys2008}%
  \BibitemOpen
  \bibfield  {author} {\bibinfo {author} {\bibfnamefont {J.~C.}\ \bibnamefont
  {Sankey}}, \bibinfo {author} {\bibfnamefont {Y.-T.}\ \bibnamefont {Cui}},
  \bibinfo {author} {\bibfnamefont {J.~Z.}\ \bibnamefont {Sun}}, \bibinfo
  {author} {\bibfnamefont {J.~C.}\ \bibnamefont {Slonczewski}}, \bibinfo
  {author} {\bibfnamefont {R.~A.}\ \bibnamefont {Buhrman}}, \ and\ \bibinfo
  {author} {\bibfnamefont {D.~C.}\ \bibnamefont {Ralph}},\ }\href@noop {}
  {\bibfield  {journal} {\bibinfo  {journal} {Nature Phys.}\ }\textbf {\bibinfo
  {volume} {4}},\ \bibinfo {pages} {67} (\bibinfo {year} {2008})}\BibitemShut
  {NoStop}%
\bibitem [{\citenamefont {Kubota}\ \emph {et~al.}(2008)\citenamefont {Kubota},
  \citenamefont {Fukushima}, \citenamefont {Yakushiji}, \citenamefont
  {Nagahama}, \citenamefont {Yuasa}, \citenamefont {Ando}, \citenamefont
  {Maehara}, \citenamefont {Nagamine}, \citenamefont {Tsunekawa}, \citenamefont
  {Djayaprawira}, \citenamefont {Watanabe},\ and\ \citenamefont
  {Suzuki}}]{KubotaNphys2008}%
  \BibitemOpen
  \bibfield  {author} {\bibinfo {author} {\bibfnamefont {H.}~\bibnamefont
  {Kubota}}, \bibinfo {author} {\bibfnamefont {A.}~\bibnamefont {Fukushima}},
  \bibinfo {author} {\bibfnamefont {K.}~\bibnamefont {Yakushiji}}, \bibinfo
  {author} {\bibfnamefont {T.}~\bibnamefont {Nagahama}}, \bibinfo {author}
  {\bibfnamefont {S.}~\bibnamefont {Yuasa}}, \bibinfo {author} {\bibfnamefont
  {K.}~\bibnamefont {Ando}}, \bibinfo {author} {\bibfnamefont {H.}~\bibnamefont
  {Maehara}}, \bibinfo {author} {\bibfnamefont {Y.}~\bibnamefont {Nagamine}},
  \bibinfo {author} {\bibfnamefont {K.}~\bibnamefont {Tsunekawa}}, \bibinfo
  {author} {\bibfnamefont {D.~D.}\ \bibnamefont {Djayaprawira}}, \bibinfo
  {author} {\bibfnamefont {N.}~\bibnamefont {Watanabe}}, \ and\ \bibinfo
  {author} {\bibfnamefont {Y.}~\bibnamefont {Suzuki}},\ }\href@noop {}
  {\bibfield  {journal} {\bibinfo  {journal} {Nature Phys.}\ }\textbf {\bibinfo
  {volume} {4}},\ \bibinfo {pages} {37} (\bibinfo {year} {2008})}\BibitemShut
  {NoStop}%
\bibitem [{\citenamefont {Chen}\ \emph {et~al.}(2008)\citenamefont {Chen},
  \citenamefont {Beaujour}, \citenamefont {de~Loubens}, \citenamefont {Kent},\
  and\ \citenamefont {Sun}}]{ChenAPL2008}%
  \BibitemOpen
  \bibfield  {author} {\bibinfo {author} {\bibfnamefont {W.}~\bibnamefont
  {Chen}}, \bibinfo {author} {\bibfnamefont {J.-M.~L.}\ \bibnamefont
  {Beaujour}}, \bibinfo {author} {\bibfnamefont {G.}~\bibnamefont
  {de~Loubens}}, \bibinfo {author} {\bibfnamefont {A.~D.}\ \bibnamefont
  {Kent}}, \ and\ \bibinfo {author} {\bibfnamefont {J.~Z.}\ \bibnamefont
  {Sun}},\ }\href@noop {} {\bibfield  {journal} {\bibinfo  {journal} {Appl.
  Phys. Lett.}\ }\textbf {\bibinfo {volume} {92}},\ \bibinfo {pages} {012507}
  (\bibinfo {year} {2008})}\BibitemShut {NoStop}%
\bibitem [{\citenamefont {Fang}\ \emph {et~al.}(2011)\citenamefont {Fang},
  \citenamefont {Kurebayashi}, \citenamefont {Wunderlich}, \citenamefont
  {V\'{y}born\'{y}}, \citenamefont {Z\^{a}rbo}, \citenamefont {Campion},
  \citenamefont {Casiraghi}, \citenamefont {Gallagher}, \citenamefont
  {Jungwirth},\ and\ \citenamefont {Ferguson}}]{FangNNano2011}%
  \BibitemOpen
  \bibfield  {author} {\bibinfo {author} {\bibfnamefont {D.}~\bibnamefont
  {Fang}}, \bibinfo {author} {\bibfnamefont {H.}~\bibnamefont {Kurebayashi}},
  \bibinfo {author} {\bibfnamefont {J.}~\bibnamefont {Wunderlich}}, \bibinfo
  {author} {\bibfnamefont {K.}~\bibnamefont {V\'{y}born\'{y}}}, \bibinfo
  {author} {\bibfnamefont {L.~P.}\ \bibnamefont {Z\^{a}rbo}}, \bibinfo {author}
  {\bibfnamefont {R.~P.}\ \bibnamefont {Campion}}, \bibinfo {author}
  {\bibfnamefont {A.}~\bibnamefont {Casiraghi}}, \bibinfo {author}
  {\bibfnamefont {B.~L.}\ \bibnamefont {Gallagher}}, \bibinfo {author}
  {\bibfnamefont {T.}~\bibnamefont {Jungwirth}}, \ and\ \bibinfo {author}
  {\bibfnamefont {A.~J.}\ \bibnamefont {Ferguson}},\ }\href@noop {} {\bibfield
  {journal} {\bibinfo  {journal} {Nature Nano.}\ }\textbf {\bibinfo {volume}
  {6}},\ \bibinfo {pages} {413} (\bibinfo {year} {2011})}\BibitemShut {NoStop}%
\bibitem [{\citenamefont {Liu}\ \emph {et~al.}(2011)\citenamefont {Liu},
  \citenamefont {Moriyama}, \citenamefont {Ralph},\ and\ \citenamefont
  {Buhrman}}]{LiuPRL2011}%
  \BibitemOpen
  \bibfield  {author} {\bibinfo {author} {\bibfnamefont {L.}~\bibnamefont
  {Liu}}, \bibinfo {author} {\bibfnamefont {T.}~\bibnamefont {Moriyama}},
  \bibinfo {author} {\bibfnamefont {D.~C.}\ \bibnamefont {Ralph}}, \ and\
  \bibinfo {author} {\bibfnamefont {R.~A.}\ \bibnamefont {Buhrman}},\ }\href
  {\doibase 10.1103/PhysRevLett.106.036601} {\bibfield  {journal} {\bibinfo
  {journal} {Phys. Rev. Lett.}\ }\textbf {\bibinfo {volume} {106}},\ \bibinfo
  {pages} {036601} (\bibinfo {year} {2011})}\BibitemShut {NoStop}%
\bibitem [{\citenamefont {Bai}\ \emph {et~al.}(2013)\citenamefont {Bai},
  \citenamefont {Hyde}, \citenamefont {Gui}, \citenamefont {Hu}, \citenamefont
  {Vlaminck}, \citenamefont {Pearson}, \citenamefont {Bader},\ and\
  \citenamefont {Hoffmann}}]{BaiPRL2013}%
  \BibitemOpen
  \bibfield  {author} {\bibinfo {author} {\bibfnamefont {L.}~\bibnamefont
  {Bai}}, \bibinfo {author} {\bibfnamefont {P.}~\bibnamefont {Hyde}}, \bibinfo
  {author} {\bibfnamefont {Y.~S.}\ \bibnamefont {Gui}}, \bibinfo {author}
  {\bibfnamefont {C.-M.}\ \bibnamefont {Hu}}, \bibinfo {author} {\bibfnamefont
  {V.}~\bibnamefont {Vlaminck}}, \bibinfo {author} {\bibfnamefont {J.~E.}\
  \bibnamefont {Pearson}}, \bibinfo {author} {\bibfnamefont {S.~D.}\
  \bibnamefont {Bader}}, \ and\ \bibinfo {author} {\bibfnamefont
  {A.}~\bibnamefont {Hoffmann}},\ }\href@noop {} {\bibfield  {journal}
  {\bibinfo  {journal} {Phys. Rev. Lett.}\ }\textbf {\bibinfo {volume} {111}},\
  \bibinfo {pages} {217602} (\bibinfo {year} {2013})}\BibitemShut {NoStop}%
\bibitem [{\citenamefont {Gui}\ \emph {et~al.}(2013)\citenamefont {Gui},
  \citenamefont {Bai},\ and\ \citenamefont {Hu}}]{GuiSChinaPMA2013}%
  \BibitemOpen
  \bibfield  {author} {\bibinfo {author} {\bibfnamefont {Y.}~\bibnamefont
  {Gui}}, \bibinfo {author} {\bibfnamefont {L.}~\bibnamefont {Bai}}, \ and\
  \bibinfo {author} {\bibfnamefont {C.-M.}\ \bibnamefont {Hu}},\ }\href@noop {}
  {\bibfield  {journal} {\bibinfo  {journal} {Sci. China Phys. Mech. Astron.}\
  }\textbf {\bibinfo {volume} {56}},\ \bibinfo {pages} {124–141} (\bibinfo
  {year} {2013})}\BibitemShut {NoStop}%
\bibitem [{\citenamefont {Mellnik}\ \emph {et~al.}(2014)\citenamefont
  {Mellnik}, \citenamefont {Lee}, \citenamefont {Richardella}, \citenamefont
  {Grab}, \citenamefont {Mintun}, \citenamefont {Fischer}, \citenamefont
  {Vaezi}, \citenamefont {Manchon}, \citenamefont {Kim}, \citenamefont
  {Samarth},\ and\ \citenamefont {Ralph}}]{MellnikNature2014}%
  \BibitemOpen
  \bibfield  {author} {\bibinfo {author} {\bibfnamefont {A.~R.}\ \bibnamefont
  {Mellnik}}, \bibinfo {author} {\bibfnamefont {J.~S.}\ \bibnamefont {Lee}},
  \bibinfo {author} {\bibfnamefont {A.}~\bibnamefont {Richardella}}, \bibinfo
  {author} {\bibfnamefont {J.~L.}\ \bibnamefont {Grab}}, \bibinfo {author}
  {\bibfnamefont {P.~J.}\ \bibnamefont {Mintun}}, \bibinfo {author}
  {\bibfnamefont {M.~H.}\ \bibnamefont {Fischer}}, \bibinfo {author}
  {\bibfnamefont {A.}~\bibnamefont {Vaezi}}, \bibinfo {author} {\bibfnamefont
  {A.}~\bibnamefont {Manchon}}, \bibinfo {author} {\bibfnamefont {E.-A.}\
  \bibnamefont {Kim}}, \bibinfo {author} {\bibfnamefont {N.}~\bibnamefont
  {Samarth}}, \ and\ \bibinfo {author} {\bibfnamefont {D.~C.}\ \bibnamefont
  {Ralph}},\ }\href@noop {} {\bibfield  {journal} {\bibinfo  {journal}
  {Nature}\ }\textbf {\bibinfo {volume} {511}},\ \bibinfo {pages} {449}
  (\bibinfo {year} {2014})}\BibitemShut {NoStop}%
\bibitem [{\citenamefont {Zhang}\ \emph
  {et~al.}(2014{\natexlab{b}})\citenamefont {Zhang}, \citenamefont
  {Jungfleisch}, \citenamefont {Jiang}, \citenamefont {Pearson}, \citenamefont
  {Hoffmann}, \citenamefont {Freimuth},\ and\ \citenamefont
  {Mokrousov}}]{ZhangWeiPRL2014}%
  \BibitemOpen
  \bibfield  {author} {\bibinfo {author} {\bibfnamefont {W.}~\bibnamefont
  {Zhang}}, \bibinfo {author} {\bibfnamefont {M.~B.}\ \bibnamefont
  {Jungfleisch}}, \bibinfo {author} {\bibfnamefont {W.}~\bibnamefont {Jiang}},
  \bibinfo {author} {\bibfnamefont {J.~E.}\ \bibnamefont {Pearson}}, \bibinfo
  {author} {\bibfnamefont {A.}~\bibnamefont {Hoffmann}}, \bibinfo {author}
  {\bibfnamefont {F.}~\bibnamefont {Freimuth}}, \ and\ \bibinfo {author}
  {\bibfnamefont {Y.}~\bibnamefont {Mokrousov}},\ }\href@noop {} {\bibfield
  {journal} {\bibinfo  {journal} {Phys. Rev. Lett.}\ }\textbf {\bibinfo
  {volume} {113}},\ \bibinfo {pages} {196602} (\bibinfo {year}
  {2014}{\natexlab{b}})}\BibitemShut {NoStop}%
\bibitem [{\citenamefont {Rojas-S\'anchez}\ \emph {et~al.}(2014)\citenamefont
  {Rojas-S\'anchez}, \citenamefont {Reyren}, \citenamefont {Laczkowski},
  \citenamefont {Savero}, \citenamefont {Attan\'e}, \citenamefont {Deranlot},
  \citenamefont {Jamet}, \citenamefont {George}, \citenamefont {Vila},\ and\
  \citenamefont {Jaffr\`es}}]{SanchezPRL2014}%
  \BibitemOpen
  \bibfield  {author} {\bibinfo {author} {\bibfnamefont {J.-C.}\ \bibnamefont
  {Rojas-S\'anchez}}, \bibinfo {author} {\bibfnamefont {N.}~\bibnamefont
  {Reyren}}, \bibinfo {author} {\bibfnamefont {P.}~\bibnamefont {Laczkowski}},
  \bibinfo {author} {\bibfnamefont {W.}~\bibnamefont {Savero}}, \bibinfo
  {author} {\bibfnamefont {J.-P.}\ \bibnamefont {Attan\'e}}, \bibinfo {author}
  {\bibfnamefont {C.}~\bibnamefont {Deranlot}}, \bibinfo {author}
  {\bibfnamefont {M.}~\bibnamefont {Jamet}}, \bibinfo {author} {\bibfnamefont
  {J.-M.}\ \bibnamefont {George}}, \bibinfo {author} {\bibfnamefont
  {L.}~\bibnamefont {Vila}}, \ and\ \bibinfo {author} {\bibfnamefont
  {H.}~\bibnamefont {Jaffr\`es}},\ }\href@noop {} {\bibfield  {journal}
  {\bibinfo  {journal} {Phys. Rev. Lett.}\ }\textbf {\bibinfo {volume} {112}},\
  \bibinfo {pages} {106602} (\bibinfo {year} {2014})}\BibitemShut {NoStop}%
\bibitem [{\citenamefont {Sklenar}\ \emph {et~al.}(2015)\citenamefont
  {Sklenar}, \citenamefont {Zhang}, \citenamefont {Jungfleisch}, \citenamefont
  {Jiang}, \citenamefont {Chang}, \citenamefont {Pearson}, \citenamefont {Wu},
  \citenamefont {Ketterson},\ and\ \citenamefont {Hoffmann}}]{SklenarPRB2015}%
  \BibitemOpen
  \bibfield  {author} {\bibinfo {author} {\bibfnamefont {J.}~\bibnamefont
  {Sklenar}}, \bibinfo {author} {\bibfnamefont {W.}~\bibnamefont {Zhang}},
  \bibinfo {author} {\bibfnamefont {M.~B.}\ \bibnamefont {Jungfleisch}},
  \bibinfo {author} {\bibfnamefont {W.}~\bibnamefont {Jiang}}, \bibinfo
  {author} {\bibfnamefont {H.}~\bibnamefont {Chang}}, \bibinfo {author}
  {\bibfnamefont {J.~E.}\ \bibnamefont {Pearson}}, \bibinfo {author}
  {\bibfnamefont {M.}~\bibnamefont {Wu}}, \bibinfo {author} {\bibfnamefont
  {J.~B.}\ \bibnamefont {Ketterson}}, \ and\ \bibinfo {author} {\bibfnamefont
  {A.}~\bibnamefont {Hoffmann}},\ }\href {\doibase 10.1103/PhysRevB.92.174406}
  {\bibfield  {journal} {\bibinfo  {journal} {Phys. Rev. B}\ }\textbf {\bibinfo
  {volume} {92}},\ \bibinfo {pages} {174406} (\bibinfo {year}
  {2015})}\BibitemShut {NoStop}%
\bibitem [{\citenamefont {Nan}\ \emph {et~al.}(2015)\citenamefont {Nan},
  \citenamefont {Emori}, \citenamefont {Boone}, \citenamefont {Wang},
  \citenamefont {Oxholm}, \citenamefont {Jones}, \citenamefont {Howe},
  \citenamefont {Brown},\ and\ \citenamefont {Sun}}]{NanPRB2015}%
  \BibitemOpen
  \bibfield  {author} {\bibinfo {author} {\bibfnamefont {T.}~\bibnamefont
  {Nan}}, \bibinfo {author} {\bibfnamefont {S.}~\bibnamefont {Emori}}, \bibinfo
  {author} {\bibfnamefont {C.~T.}\ \bibnamefont {Boone}}, \bibinfo {author}
  {\bibfnamefont {X.}~\bibnamefont {Wang}}, \bibinfo {author} {\bibfnamefont
  {T.~M.}\ \bibnamefont {Oxholm}}, \bibinfo {author} {\bibfnamefont {J.~G.}\
  \bibnamefont {Jones}}, \bibinfo {author} {\bibfnamefont {B.~M.}\ \bibnamefont
  {Howe}}, \bibinfo {author} {\bibfnamefont {G.~J.}\ \bibnamefont {Brown}}, \
  and\ \bibinfo {author} {\bibfnamefont {N.~X.}\ \bibnamefont {Sun}},\ }\href
  {\doibase 10.1103/PhysRevB.91.214416} {\bibfield  {journal} {\bibinfo
  {journal} {Phys. Rev. B}\ }\textbf {\bibinfo {volume} {91}},\ \bibinfo
  {pages} {214416} (\bibinfo {year} {2015})}\BibitemShut {NoStop}%
\bibitem [{\citenamefont {Jungfleisch}\ \emph {et~al.}(2016)\citenamefont
  {Jungfleisch}, \citenamefont {Zhang}, \citenamefont {Sklenar}, \citenamefont
  {Ding}, \citenamefont {Jiang}, \citenamefont {Chang}, \citenamefont {Fradin},
  \citenamefont {Pearson}, \citenamefont {Ketterson}, \citenamefont {Novosad},
  \citenamefont {Wu},\ and\ \citenamefont {Hoffmann}}]{JungfleischPRL2016}%
  \BibitemOpen
  \bibfield  {author} {\bibinfo {author} {\bibfnamefont {M.~B.}\ \bibnamefont
  {Jungfleisch}}, \bibinfo {author} {\bibfnamefont {W.}~\bibnamefont {Zhang}},
  \bibinfo {author} {\bibfnamefont {J.}~\bibnamefont {Sklenar}}, \bibinfo
  {author} {\bibfnamefont {J.}~\bibnamefont {Ding}}, \bibinfo {author}
  {\bibfnamefont {W.}~\bibnamefont {Jiang}}, \bibinfo {author} {\bibfnamefont
  {H.}~\bibnamefont {Chang}}, \bibinfo {author} {\bibfnamefont {F.~Y.}\
  \bibnamefont {Fradin}}, \bibinfo {author} {\bibfnamefont {J.~E.}\
  \bibnamefont {Pearson}}, \bibinfo {author} {\bibfnamefont {J.~B.}\
  \bibnamefont {Ketterson}}, \bibinfo {author} {\bibfnamefont {V.}~\bibnamefont
  {Novosad}}, \bibinfo {author} {\bibfnamefont {M.}~\bibnamefont {Wu}}, \ and\
  \bibinfo {author} {\bibfnamefont {A.}~\bibnamefont {Hoffmann}},\ }\href
  {\doibase 10.1103/PhysRevLett.116.057601} {\bibfield  {journal} {\bibinfo
  {journal} {Phys. Rev. Lett.}\ }\textbf {\bibinfo {volume} {116}},\ \bibinfo
  {pages} {057601} (\bibinfo {year} {2016})}\BibitemShut {NoStop}%
\bibitem [{\citenamefont {Harder}\ \emph {et~al.}(2016)\citenamefont {Harder},
  \citenamefont {Gui},\ and\ \citenamefont {Hu}}]{HarderPhysRep2016}%
  \BibitemOpen
  \bibfield  {author} {\bibinfo {author} {\bibfnamefont {M.}~\bibnamefont
  {Harder}}, \bibinfo {author} {\bibfnamefont {Y.}~\bibnamefont {Gui}}, \ and\
  \bibinfo {author} {\bibfnamefont {C.-M.}\ \bibnamefont {Hu}},\ }\href@noop {}
  {\bibfield  {journal} {\bibinfo  {journal} {Phys. Rep.}\ }\textbf {\bibinfo
  {volume} {661}},\ \bibinfo {pages} {1} (\bibinfo {year} {2016})}\BibitemShut
  {NoStop}%
\bibitem [{\citenamefont {Li}\ \emph {et~al.}(2016)\citenamefont {Li},
  \citenamefont {Zhang}, \citenamefont {Ding}, \citenamefont {Pearson},
  \citenamefont {Novosad},\ and\ \citenamefont {Hoffmann}}]{LiNanoscale2016}%
  \BibitemOpen
  \bibfield  {author} {\bibinfo {author} {\bibfnamefont {S.}~\bibnamefont
  {Li}}, \bibinfo {author} {\bibfnamefont {W.}~\bibnamefont {Zhang}}, \bibinfo
  {author} {\bibfnamefont {J.}~\bibnamefont {Ding}}, \bibinfo {author}
  {\bibfnamefont {J.~E.}\ \bibnamefont {Pearson}}, \bibinfo {author}
  {\bibfnamefont {V.}~\bibnamefont {Novosad}}, \ and\ \bibinfo {author}
  {\bibfnamefont {A.}~\bibnamefont {Hoffmann}},\ }\href@noop {} {\bibfield
  {journal} {\bibinfo  {journal} {Nanoscale}\ }\textbf {\bibinfo {volume}
  {8}},\ \bibinfo {pages} {388} (\bibinfo {year} {2016})}\BibitemShut {NoStop}%
\bibitem [{\citenamefont {Hyde}\ \emph {et~al.}(2014)\citenamefont {Hyde},
  \citenamefont {Bai}, \citenamefont {Kumar}, \citenamefont {Southern},
  \citenamefont {Hu}, \citenamefont {Huang}, \citenamefont {Miao},\ and\
  \citenamefont {Chien}}]{HydePRB2014}%
  \BibitemOpen
  \bibfield  {author} {\bibinfo {author} {\bibfnamefont {P.}~\bibnamefont
  {Hyde}}, \bibinfo {author} {\bibfnamefont {L.}~\bibnamefont {Bai}}, \bibinfo
  {author} {\bibfnamefont {D.~M.~J.}\ \bibnamefont {Kumar}}, \bibinfo {author}
  {\bibfnamefont {B.~W.}\ \bibnamefont {Southern}}, \bibinfo {author}
  {\bibfnamefont {C.-M.}\ \bibnamefont {Hu}}, \bibinfo {author} {\bibfnamefont
  {S.~Y.}\ \bibnamefont {Huang}}, \bibinfo {author} {\bibfnamefont {B.~F.}\
  \bibnamefont {Miao}}, \ and\ \bibinfo {author} {\bibfnamefont {C.~L.}\
  \bibnamefont {Chien}},\ }\href {\doibase 10.1103/PhysRevB.89.180404}
  {\bibfield  {journal} {\bibinfo  {journal} {Phys. Rev. B}\ }\textbf {\bibinfo
  {volume} {89}},\ \bibinfo {pages} {180404} (\bibinfo {year}
  {2014})}\BibitemShut {NoStop}%
\bibitem [{\citenamefont {Yang}\ \emph {et~al.}(2020)\citenamefont {Yang},
  \citenamefont {Wei}, \citenamefont {Wan}, \citenamefont {Xing}, \citenamefont
  {Yan}, \citenamefont {Wang}, \citenamefont {Fang}, \citenamefont {Guo},
  \citenamefont {Yu},\ and\ \citenamefont {Han}}]{YangPRB2020}%
  \BibitemOpen
  \bibfield  {author} {\bibinfo {author} {\bibfnamefont {W.~L.}\ \bibnamefont
  {Yang}}, \bibinfo {author} {\bibfnamefont {J.~W.}\ \bibnamefont {Wei}},
  \bibinfo {author} {\bibfnamefont {C.~H.}\ \bibnamefont {Wan}}, \bibinfo
  {author} {\bibfnamefont {Y.~W.}\ \bibnamefont {Xing}}, \bibinfo {author}
  {\bibfnamefont {Z.~R.}\ \bibnamefont {Yan}}, \bibinfo {author} {\bibfnamefont
  {X.}~\bibnamefont {Wang}}, \bibinfo {author} {\bibfnamefont {C.}~\bibnamefont
  {Fang}}, \bibinfo {author} {\bibfnamefont {C.~Y.}\ \bibnamefont {Guo}},
  \bibinfo {author} {\bibfnamefont {G.~Q.}\ \bibnamefont {Yu}}, \ and\ \bibinfo
  {author} {\bibfnamefont {X.~F.}\ \bibnamefont {Han}},\ }\href {\doibase
  10.1103/PhysRevB.101.064412} {\bibfield  {journal} {\bibinfo  {journal}
  {Phys. Rev. B}\ }\textbf {\bibinfo {volume} {101}},\ \bibinfo {pages}
  {064412} (\bibinfo {year} {2020})}\BibitemShut {NoStop}%
\bibitem [{\citenamefont {Li}\ \emph {et~al.}(2019{\natexlab{b}})\citenamefont
  {Li}, \citenamefont {Zeng}, \citenamefont {Zhang}, \citenamefont {Shin},
  \citenamefont {Saglam}, \citenamefont {Karakas}, \citenamefont {Ozatay},
  \citenamefont {Pearson}, \citenamefont {Heinonen}, \citenamefont {Wu},
  \citenamefont {Hoffmann},\ and\ \citenamefont {Zhang}}]{LiPRL2019_CoFe}%
  \BibitemOpen
  \bibfield  {author} {\bibinfo {author} {\bibfnamefont {Y.}~\bibnamefont
  {Li}}, \bibinfo {author} {\bibfnamefont {F.}~\bibnamefont {Zeng}}, \bibinfo
  {author} {\bibfnamefont {S.~S.-L.}\ \bibnamefont {Zhang}}, \bibinfo {author}
  {\bibfnamefont {H.}~\bibnamefont {Shin}}, \bibinfo {author} {\bibfnamefont
  {H.}~\bibnamefont {Saglam}}, \bibinfo {author} {\bibfnamefont
  {V.}~\bibnamefont {Karakas}}, \bibinfo {author} {\bibfnamefont
  {O.}~\bibnamefont {Ozatay}}, \bibinfo {author} {\bibfnamefont {J.~E.}\
  \bibnamefont {Pearson}}, \bibinfo {author} {\bibfnamefont {O.~G.}\
  \bibnamefont {Heinonen}}, \bibinfo {author} {\bibfnamefont {Y.}~\bibnamefont
  {Wu}}, \bibinfo {author} {\bibfnamefont {A.}~\bibnamefont {Hoffmann}}, \ and\
  \bibinfo {author} {\bibfnamefont {W.}~\bibnamefont {Zhang}},\ }\href@noop {}
  {\bibfield  {journal} {\bibinfo  {journal} {Phys. Rev. Lett.}\ }\textbf
  {\bibinfo {volume} {122}},\ \bibinfo {pages} {117203} (\bibinfo {year}
  {2019}{\natexlab{b}})}\BibitemShut {NoStop}%
\bibitem [{\citenamefont {Hisatomi}\ \emph {et~al.}(2016)\citenamefont
  {Hisatomi}, \citenamefont {Osada}, \citenamefont {Tabuchi}, \citenamefont
  {Ishikawa}, \citenamefont {Noguchi}, \citenamefont {Yamazaki}, \citenamefont
  {Usami},\ and\ \citenamefont {Nakamura}}]{HisatomiPRB2016}%
  \BibitemOpen
  \bibfield  {author} {\bibinfo {author} {\bibfnamefont {R.}~\bibnamefont
  {Hisatomi}}, \bibinfo {author} {\bibfnamefont {A.}~\bibnamefont {Osada}},
  \bibinfo {author} {\bibfnamefont {Y.}~\bibnamefont {Tabuchi}}, \bibinfo
  {author} {\bibfnamefont {T.}~\bibnamefont {Ishikawa}}, \bibinfo {author}
  {\bibfnamefont {A.}~\bibnamefont {Noguchi}}, \bibinfo {author} {\bibfnamefont
  {R.}~\bibnamefont {Yamazaki}}, \bibinfo {author} {\bibfnamefont
  {K.}~\bibnamefont {Usami}}, \ and\ \bibinfo {author} {\bibfnamefont
  {Y.}~\bibnamefont {Nakamura}},\ }\href {\doibase 10.1103/PhysRevB.93.174427}
  {\bibfield  {journal} {\bibinfo  {journal} {Phys. Rev. B}\ }\textbf {\bibinfo
  {volume} {93}},\ \bibinfo {pages} {174427} (\bibinfo {year}
  {2016})}\BibitemShut {NoStop}%
\bibitem [{\citenamefont {Yoon}\ \emph {et~al.}(2016)\citenamefont {Yoon},
  \citenamefont {Liu},\ and\ \citenamefont {McMichael}}]{YoonPRB2016}%
  \BibitemOpen
  \bibfield  {author} {\bibinfo {author} {\bibfnamefont {S.}~\bibnamefont
  {Yoon}}, \bibinfo {author} {\bibfnamefont {J.}~\bibnamefont {Liu}}, \ and\
  \bibinfo {author} {\bibfnamefont {R.~D.}\ \bibnamefont {McMichael}},\ }\href
  {\doibase 10.1103/PhysRevB.93.144423} {\bibfield  {journal} {\bibinfo
  {journal} {Phys. Rev. B}\ }\textbf {\bibinfo {volume} {93}},\ \bibinfo
  {pages} {144423} (\bibinfo {year} {2016})}\BibitemShut {NoStop}%
\bibitem [{\citenamefont {Capua}\ \emph {et~al.}(2017)\citenamefont {Capua},
  \citenamefont {Wang}, \citenamefont {Yang}, \citenamefont {Rettner},
  \citenamefont {Phung},\ and\ \citenamefont {Parkin}}]{CapuaPRB2017}%
  \BibitemOpen
  \bibfield  {author} {\bibinfo {author} {\bibfnamefont {A.}~\bibnamefont
  {Capua}}, \bibinfo {author} {\bibfnamefont {T.}~\bibnamefont {Wang}},
  \bibinfo {author} {\bibfnamefont {S.-H.}\ \bibnamefont {Yang}}, \bibinfo
  {author} {\bibfnamefont {C.}~\bibnamefont {Rettner}}, \bibinfo {author}
  {\bibfnamefont {T.}~\bibnamefont {Phung}}, \ and\ \bibinfo {author}
  {\bibfnamefont {S.~S.~P.}\ \bibnamefont {Parkin}},\ }\href {\doibase
  10.1103/PhysRevB.95.064401} {\bibfield  {journal} {\bibinfo  {journal} {Phys.
  Rev. B}\ }\textbf {\bibinfo {volume} {95}},\ \bibinfo {pages} {064401}
  (\bibinfo {year} {2017})}\BibitemShut {NoStop}%
\bibitem [{\citenamefont {Li}\ \emph {et~al.}(2019{\natexlab{c}})\citenamefont
  {Li}, \citenamefont {Saglam}, \citenamefont {Zhang}, \citenamefont
  {Bidthanapally}, \citenamefont {Xiong}, \citenamefont {Pearson},
  \citenamefont {Novosad}, \citenamefont {Qu}, \citenamefont {Srinivasan},
  \citenamefont {Hoffmann},\ and\ \citenamefont {Zhang}}]{LiPRApplied2019}%
  \BibitemOpen
  \bibfield  {author} {\bibinfo {author} {\bibfnamefont {Y.}~\bibnamefont
  {Li}}, \bibinfo {author} {\bibfnamefont {H.}~\bibnamefont {Saglam}}, \bibinfo
  {author} {\bibfnamefont {Z.}~\bibnamefont {Zhang}}, \bibinfo {author}
  {\bibfnamefont {R.}~\bibnamefont {Bidthanapally}}, \bibinfo {author}
  {\bibfnamefont {Y.}~\bibnamefont {Xiong}}, \bibinfo {author} {\bibfnamefont
  {J.~E.}\ \bibnamefont {Pearson}}, \bibinfo {author} {\bibfnamefont
  {V.}~\bibnamefont {Novosad}}, \bibinfo {author} {\bibfnamefont
  {H.}~\bibnamefont {Qu}}, \bibinfo {author} {\bibfnamefont {G.}~\bibnamefont
  {Srinivasan}}, \bibinfo {author} {\bibfnamefont {A.}~\bibnamefont
  {Hoffmann}}, \ and\ \bibinfo {author} {\bibfnamefont {W.}~\bibnamefont
  {Zhang}},\ }\href {\doibase 10.1103/PhysRevApplied.11.034047} {\bibfield
  {journal} {\bibinfo  {journal} {Phys. Rev. Applied}\ }\textbf {\bibinfo
  {volume} {11}},\ \bibinfo {pages} {034047} (\bibinfo {year}
  {2019}{\natexlab{c}})}\BibitemShut {NoStop}%
\bibitem [{\citenamefont {Li}\ \emph {et~al.}(2019{\natexlab{d}})\citenamefont
  {Li}, \citenamefont {Zeng}, \citenamefont {Saglam}, \citenamefont {Sklenar},
  \citenamefont {Pearson}, \citenamefont {Sebastian}, \citenamefont {Wu},
  \citenamefont {Hoffmann},\ and\ \citenamefont {Zhang}}]{LiIEEETransMagn2019}%
  \BibitemOpen
  \bibfield  {author} {\bibinfo {author} {\bibfnamefont {Y.}~\bibnamefont
  {Li}}, \bibinfo {author} {\bibfnamefont {F.}~\bibnamefont {Zeng}}, \bibinfo
  {author} {\bibfnamefont {H.}~\bibnamefont {Saglam}}, \bibinfo {author}
  {\bibfnamefont {J.}~\bibnamefont {Sklenar}}, \bibinfo {author} {\bibfnamefont
  {J.~E.}\ \bibnamefont {Pearson}}, \bibinfo {author} {\bibfnamefont
  {T.}~\bibnamefont {Sebastian}}, \bibinfo {author} {\bibfnamefont
  {Y.}~\bibnamefont {Wu}}, \bibinfo {author} {\bibfnamefont {A.}~\bibnamefont
  {Hoffmann}}, \ and\ \bibinfo {author} {\bibfnamefont {W.}~\bibnamefont
  {Zhang}},\ }\href@noop {} {\bibfield  {journal} {\bibinfo  {journal} {IEEE
  Trans. Magn.}\ }\textbf {\bibinfo {volume} {55}},\ \bibinfo {pages} {6100605}
  (\bibinfo {year} {2019}{\natexlab{d}})}\BibitemShut {NoStop}%
\bibitem [{\citenamefont {Wei}\ \emph {et~al.}(2020)\citenamefont {Wei},
  \citenamefont {He}, \citenamefont {Wang}, \citenamefont {Xu}, \citenamefont
  {Liu}, \citenamefont {Guang}, \citenamefont {Wan}, \citenamefont {Feng},
  \citenamefont {Yu},\ and\ \citenamefont {Han}}]{WeiJinwuPRApplied2020}%
  \BibitemOpen
  \bibfield  {author} {\bibinfo {author} {\bibfnamefont {J.}~\bibnamefont
  {Wei}}, \bibinfo {author} {\bibfnamefont {C.}~\bibnamefont {He}}, \bibinfo
  {author} {\bibfnamefont {X.}~\bibnamefont {Wang}}, \bibinfo {author}
  {\bibfnamefont {H.}~\bibnamefont {Xu}}, \bibinfo {author} {\bibfnamefont
  {Y.}~\bibnamefont {Liu}}, \bibinfo {author} {\bibfnamefont {Y.}~\bibnamefont
  {Guang}}, \bibinfo {author} {\bibfnamefont {C.}~\bibnamefont {Wan}}, \bibinfo
  {author} {\bibfnamefont {J.}~\bibnamefont {Feng}}, \bibinfo {author}
  {\bibfnamefont {G.}~\bibnamefont {Yu}}, \ and\ \bibinfo {author}
  {\bibfnamefont {X.}~\bibnamefont {Han}},\ }\href {\doibase
  10.1103/PhysRevApplied.13.034041} {\bibfield  {journal} {\bibinfo  {journal}
  {Phys. Rev. Applied}\ }\textbf {\bibinfo {volume} {13}},\ \bibinfo {pages}
  {034041} (\bibinfo {year} {2020})}\BibitemShut {NoStop}%
\bibitem [{\citenamefont {Wang}\ \emph
  {et~al.}(2019{\natexlab{b}})\citenamefont {Wang}, \citenamefont {Wang},
  \citenamefont {Amin}, \citenamefont {Wang}, \citenamefont {Radhakrishnan},
  \citenamefont {Davidson}, \citenamefont {Allen}, \citenamefont {Silva},
  \citenamefont {Ohldag}, \citenamefont {Balzar}, \citenamefont {Zink},
  \citenamefont {Haney}, \citenamefont {Xiao}, \citenamefont {Cahill},
  \citenamefont {Lorenz},\ and\ \citenamefont {Fan}}]{WangNNano2019}%
  \BibitemOpen
  \bibfield  {author} {\bibinfo {author} {\bibfnamefont {W.}~\bibnamefont
  {Wang}}, \bibinfo {author} {\bibfnamefont {T.}~\bibnamefont {Wang}}, \bibinfo
  {author} {\bibfnamefont {V.~P.}\ \bibnamefont {Amin}}, \bibinfo {author}
  {\bibfnamefont {Y.}~\bibnamefont {Wang}}, \bibinfo {author} {\bibfnamefont
  {A.}~\bibnamefont {Radhakrishnan}}, \bibinfo {author} {\bibfnamefont
  {A.}~\bibnamefont {Davidson}}, \bibinfo {author} {\bibfnamefont {S.~R.}\
  \bibnamefont {Allen}}, \bibinfo {author} {\bibfnamefont {T.~J.}\ \bibnamefont
  {Silva}}, \bibinfo {author} {\bibfnamefont {H.}~\bibnamefont {Ohldag}},
  \bibinfo {author} {\bibfnamefont {D.}~\bibnamefont {Balzar}}, \bibinfo
  {author} {\bibfnamefont {B.~L.}\ \bibnamefont {Zink}}, \bibinfo {author}
  {\bibfnamefont {P.~M.}\ \bibnamefont {Haney}}, \bibinfo {author}
  {\bibfnamefont {J.~Q.}\ \bibnamefont {Xiao}}, \bibinfo {author}
  {\bibfnamefont {D.~G.}\ \bibnamefont {Cahill}}, \bibinfo {author}
  {\bibfnamefont {V.~O.}\ \bibnamefont {Lorenz}}, \ and\ \bibinfo {author}
  {\bibfnamefont {X.}~\bibnamefont {Fan}},\ }\href@noop {} {\bibfield
  {journal} {\bibinfo  {journal} {Nature Nano.}\ }\textbf {\bibinfo {volume}
  {14}},\ \bibinfo {pages} {819} (\bibinfo {year}
  {2019}{\natexlab{b}})}\BibitemShut {NoStop}%
\bibitem [{\citenamefont {Chiba}\ \emph {et~al.}(2014)\citenamefont {Chiba},
  \citenamefont {Bauer},\ and\ \citenamefont {Takahashi}}]{ChibaPRApplied2014}%
  \BibitemOpen
  \bibfield  {author} {\bibinfo {author} {\bibfnamefont {T.}~\bibnamefont
  {Chiba}}, \bibinfo {author} {\bibfnamefont {G.~E.~W.}\ \bibnamefont {Bauer}},
  \ and\ \bibinfo {author} {\bibfnamefont {S.}~\bibnamefont {Takahashi}},\
  }\href@noop {} {\bibfield  {journal} {\bibinfo  {journal} {Phys. Rev.
  Applied}\ }\textbf {\bibinfo {volume} {2}},\ \bibinfo {pages} {034003}
  (\bibinfo {year} {2014})}\BibitemShut {NoStop}%
\bibitem [{\citenamefont {Bhoi}\ and\ \citenamefont {Kim}(2019)}]{BhoiSSP2019}%
  \BibitemOpen
  \bibfield  {author} {\bibinfo {author} {\bibfnamefont {B.}~\bibnamefont
  {Bhoi}}\ and\ \bibinfo {author} {\bibfnamefont {S.-K.}\ \bibnamefont {Kim}},\
  }\href@noop {} {\bibfield  {journal} {\bibinfo  {journal} {Solid State
  Phys.}\ }\textbf {\bibinfo {volume} {70}},\ \bibinfo {pages} {1} (\bibinfo
  {year} {2019})}\BibitemShut {NoStop}%
\bibitem [{\citenamefont {Lachance-Quirion}\ \emph {et~al.}(2019)\citenamefont
  {Lachance-Quirion}, \citenamefont {Tabuchi}, \citenamefont {Gloppe},
  \citenamefont {Usami},\ and\ \citenamefont
  {Nakamura}}]{LachanceQuirionAPEx2019}%
  \BibitemOpen
  \bibfield  {author} {\bibinfo {author} {\bibfnamefont {D.}~\bibnamefont
  {Lachance-Quirion}}, \bibinfo {author} {\bibfnamefont {Y.}~\bibnamefont
  {Tabuchi}}, \bibinfo {author} {\bibfnamefont {A.}~\bibnamefont {Gloppe}},
  \bibinfo {author} {\bibfnamefont {K.}~\bibnamefont {Usami}}, \ and\ \bibinfo
  {author} {\bibfnamefont {Y.}~\bibnamefont {Nakamura}},\ }\href@noop {}
  {\bibfield  {journal} {\bibinfo  {journal} {Appl. Phys. Express}\ }\textbf
  {\bibinfo {volume} {12}},\ \bibinfo {pages} {070101} (\bibinfo {year}
  {2019})}\BibitemShut {NoStop}%
\bibitem [{\citenamefont {Kusminskiy}()}]{KusminskiyarXiv2019}%
  \BibitemOpen
  \bibfield  {author} {\bibinfo {author} {\bibfnamefont {S.~V.}\ \bibnamefont
  {Kusminskiy}},\ }\href@noop {} {\bibfield  {journal} {\bibinfo  {journal}
  {arXiv}\ }}\bibinfo {note} {1911.11104}\BibitemShut {NoStop}%
\bibitem [{\citenamefont {Wang}\ and\ \citenamefont {Hu}(2020)}]{WangJAP2020}%
  \BibitemOpen
  \bibfield  {author} {\bibinfo {author} {\bibfnamefont {Y.-P.}\ \bibnamefont
  {Wang}}\ and\ \bibinfo {author} {\bibfnamefont {C.-M.}\ \bibnamefont {Hu}},\
  }\href@noop {} {\bibfield  {journal} {\bibinfo  {journal} {J. Appl. Phys.}\
  }\textbf {\bibinfo {volume} {127}},\ \bibinfo {pages} {130901} (\bibinfo
  {year} {2020})}\BibitemShut {NoStop}%
\bibitem [{\citenamefont {Li}\ \emph {et~al.}(2020{\natexlab{b}})\citenamefont
  {Li}, \citenamefont {Zhang}, \citenamefont {Tyberkevych}, \citenamefont
  {Kwok}, \citenamefont {Hoffmann},\ and\ \citenamefont {Novosad}}]{LiJAP2020}%
  \BibitemOpen
  \bibfield  {author} {\bibinfo {author} {\bibfnamefont {Y.}~\bibnamefont
  {Li}}, \bibinfo {author} {\bibfnamefont {W.}~\bibnamefont {Zhang}}, \bibinfo
  {author} {\bibfnamefont {V.}~\bibnamefont {Tyberkevych}}, \bibinfo {author}
  {\bibfnamefont {W.-K.}\ \bibnamefont {Kwok}}, \bibinfo {author}
  {\bibfnamefont {A.}~\bibnamefont {Hoffmann}}, \ and\ \bibinfo {author}
  {\bibfnamefont {V.}~\bibnamefont {Novosad}},\ }\href@noop {} {\bibfield
  {journal} {\bibinfo  {journal} {J. Appl. Phys.}\ }\textbf {\bibinfo {volume}
  {128}},\ \bibinfo {pages} {130902} (\bibinfo {year}
  {2020}{\natexlab{b}})}\BibitemShut {NoStop}%
\end{thebibliography}

%

\end{document}